\documentclass[aps,pra,twocolumn,showpacs,superscriptaddress]{revtex4}
\usepackage{graphicx}
\usepackage{amsmath}
\usepackage{amssymb}
\usepackage{lscape}
\usepackage{multirow}
\usepackage{longtable}

\input{epsf}

\begin{document}

\title{Scattering framework for two particles with isotropic spin-orbit coupling applicable to all energies}

\author{Q. Guan}
\affiliation{Department of Physics and Astronomy,
Washington State University,
  Pullman, Washington 99164-2814, USA}

\author{D. Blume}
\affiliation{Department of Physics and Astronomy,
Washington State University,
  Pullman, Washington 99164-2814, USA}

\date{\today}

\begin{abstract}
Previous work developed
a K-matrix formalism applicable to positive energies
for the scattering between two $s$-wave interacting particles
with two internal states, isotropic spin-orbit coupling
and vanishing center-of-mass momentum
[H. Duan, L. You and B. Gao,
Phys. Rev A {\bf{87}}, 052708 (2013)].
This work extends the formalism to the entire energy
regime. Explicit solutions are obtained
for the total angular momentum $J=0$ and $1$ channels.
The behavior of the partial cross 
sections in the negative energy regime
is analyzed in detail. 
We find that
the leading contributions
to the partial cross sections
at the negative energy thresholds 
are
governed by the spin-orbit coupling strength $k_{\text{so}}$ 
and the mass ratio.
The fact that these contributions are independent of
the two-body scattering length $a_s$ is a direct consequence
of the effective reduction of the dimensionality, and hence
of the density of states, near the scattering thresholds due to the 
single-particle spin-orbit coupling terms.
The results are analytically continued to the energy regime where
bound states exist.
It is shown that our results are consistent with results
obtained by 
alternative
approaches.
Our formulation, which can be regarded as an extension of
the standard textbook partial wave decomposition, 
can be generalized to two-body systems with 
other types of spin-orbit coupling, including 
cases where the center-of-mass momentum does not vanish.

\end{abstract}

\maketitle

\section{introduction}
\label{sec1}
Spin-momentum coupling, which is associated with the 
presence of non-Abelian gauge fields,
is crucial for a range of interesting effects 
in condensed matter physics.
Throughout this article,
we follow established terminology and refer to the coupling between 
a particle's spin degrees of freedom and its canonical
momentum as spin-orbit coupling~\cite{condense_matter,condense_matter1}.
Some of the interest in these spin-orbit coupled systems 
stems from the fact that the single-particle dispersion
curve displays Dirac rather than Schr\"odinger 
equation-type characteristics.
The realization of 
synthetic gauge fields for neutral cold atom systems
provides opportunities to
(i) mimic condensed matter phenomena and (ii)
look for novel physics not accessible with
conventional condensed matter systems.
In cold atom systems, a variety of techniques 
have been developed to create synthetic gauge fields,
including lattice shaking~\cite{sengstock} and Raman coupling~\cite{spielman}.
Raman
laser coupling schemes have already led
to the
experimental 
realization of
one-dimensional spin-orbit coupling 
(equal mixture of Rashba and 
Dresselhaus spin-orbit coupling)~\cite{spielman, zwierlein, zhang, pan}
and 
two-dimensional spin-orbit coupling~\cite{2D_SOC}.
This paper considers isotropic three-dimensional spin-orbit
coupling. 
While this type of spin-orbit coupling has not yet been
realized experimentally in cold atom systems,
several
proposals 
exist toward its experimental
realization~\cite{proposal_1, proposal_2, proposal_3}. 
The two-particle scattering framework developed in our work
for systems with short-range interactions is related to
scattering works for electronic systems with
spin-orbit coupling.
In the context of electronic systems,
the negative energy regime, which is the focus of our work,
has not received as much attention as the positive
energy regime~\cite{rashba_billiard,Novikov,Joel}. 
Thus, we expect our developments to
not only be of interest to the cold atom community but also 
to the condensed matter community.

Spin-orbit coupled cold atom systems are currently of great interest
to experimentalists and theorists.
To date a variety of exciting single-particle based phenomena
such as
Landau-Zener transitions~\cite{landau_zener},
Zitterbewegung oscillations~\cite{Zitterbewegung},
and spin wave dynamics~\cite{spin_wave}
have been studied.
Two-body interactions add a new degree of freedom to the system.
For spin-orbit coupled systems, 
unlike in the alkalis, 
the singlet and triplet channels are strongly coupled, 
giving rise to changes of 
the two-body binding energy and
the crossover 
physics in Fermi gases~\cite{H_Hu, ming_gong, crossover_hui, bound_shenoy, crossover_shenoy, rashbon_shenoy, L_Han}.
While the experimental study of these effects is still in its infancy,
first radio-frequency studies of weakly-bound
Feshbach molecules reveal that the spin-orbit coupling
terms have an appreciable effect~\cite{zhang_jing}.
In a different experiment,
the mixing of different partial waves was demonstrated explicitly
in bosonic systems with
one-dimensional spin-orbit coupling~\cite{scattering_experiment}.
For a Bose-Einstein condensate, 
the existence of a stripe phase has been predicted theoretically 
based on the mean-field Gross-Pitaevskii equation.
This new phase arises for certain Raman coupling strengths
if the
interspecies and intraspecies scattering lengths
differ~\cite{stripe_hui, jason, stripe, stringari}.
The interplay between the 
two-body interactions and the single-particle
spin-orbit coupling terms also leads to interesting few-body
effects. For example, effectively one-dimensional systems
with spin-orbit coupling allow for the realization of
spin-chain models~\cite{spin_chain_cui, spin_chain_guan}.
Moreover, Borromean three-body states 
have been predicted to exist~\cite{three_body_hui, three_body_yi, three_body_cui}.

Motivated by the developments presented in Refs.~\cite{Gao,chris}
for positive energies,
this paper develops a scattering formalism
for two particles with isotropic spin-orbit coupling
applicable to the entire
energy regime. The formulation can be regarded 
as a generalization of the usual partial wave decomposition
for two particles that are, at large interparticle
distances, fully determined by the kinetic energy.
In the presence of spin-orbit coupling, the particles'
behavior at large distances is governed by the combination
of the kinetic energy and the spin-orbit coupling term.
The presence of the spin-orbit coupling modifies
the asymptotic 
form of the wave function 
to be matched to.
Our formalism is illustrated for 
the contact $s$-wave interaction potential, 
which allows for the derivation
of analytical expressions. 
While some of the results for the $(J,M_J)=(0,0)$
channel had been derived previously~\cite{bound_shenoy, pengzhang, pengzhang1, zhenhua, xiaoling, Gao}, 
the results for the $(J,M_J)=(1,M_J)$ channel
are, to the best of our knowledge, new. 
It is shown
that the mass ratio can be used to tune the scattering 
properties.
This finding suggests rich physics for unequal-mass 
systems with spin-orbit coupling.
We find that the leading
terms of the partial cross sections are independent of 
the $s$-wave scattering length for all negative
energy scattering thresholds.
This $s$-wave scattering length
independence suggests a new type of universality,
namely a regime where the two-body scattering cross sections
are determined by the single-particle spin-orbit coupling parameter 
$k_{\text{so}}$. The effect can be traced back to
an effective reduction of the dimensionality due to the
spin-orbit coupling. While this effective dimensionality reduction
is well known and appreciated~\cite{bound_shenoy, xiaoling, review_hui}, 
its impact on threshold
laws has, to the best of our knowledge,
not been discussed in detail in the literature.
Our results for the $(J,M_J)=(0,0)$
and $(1,M_J)$ channels are related to results
obtained by alternative 
approaches~\cite{xiaoling, pengzhang, pengzhang1, zhenhua}.
It is demonstrated 
that the asymptotic
basis chosen in our work and in 
Ref.~\cite{xiaoling}
are different. With a proper unitary transformation,
the solutions can, however, be transformed into each other (see also Ref.~\cite{zhenhua}).
While we, naturally, prefer our approach, it is argued
that the use of the alternative asymptotic basis provides
a useful complementary viewpoint.
Last, our formulation provides the basis for numerical
coupled-channel calculations for systems with spin-orbit
coupling. While it was already pointed out in Ref.~\cite{Gao}
that the partial wave decomposition approach only requires
minor modifications of a typical coupled-channel code,
it is our work that shows how to set such calculations up
consistently
over the entire energy regime.

The remainder of this paper is organized as follows.
Section~\ref{sec2} introduces the general scattering framework.
This framework is then applied to the $(J,M_J)=(0,0)$
channel in Sec.~\ref{sec3}
and
to the $(J,M_J)=(1,M_J)$
channel in Sec.~\ref{sec4}.
Last, Sec.~\ref{sec5} provides a summary and an outlook.

\section{General formalism}
\label{sec2}
For two particles interacting
through a
two-body short-range interaction potential $\hat{V}_{\text{2b}}(\mathbf{r}_1-\mathbf{r}_2)$
with isotropic spin-orbit coupling terms that
are proportional to $k_{\text{so}}$,
the system Hamiltonian $\hat{H}_{\text{tot}}$ reads
\begin{eqnarray}
\label{hamiltonian_tot}
\hat{H}_{\text{tot}}&=&
\left(\frac{\hat{\mathbf{p}}_1^2}{2m_1}+\frac{\mathbf{\hat{p}}_2^2}{2m_2} \right)I_1 \otimes I_2
\nonumber \\
&+&\frac{\hbar k_{\text{so}}}{m_1}
\left(
\mathbf{\hat{p}}_1\cdot\hat{\pmb{\sigma}}_1 \right)
\otimes I_2+
\frac{\hbar k_{\text{so}}}{m_2}
I_1 \otimes \left( \mathbf{\hat{p}}_2\cdot\hat{\pmb{\sigma}}_2
\right) 
 \nonumber \\
&+& \hat{V}_{\text{2b}}(\mathbf{r}_1-\mathbf{r}_2)I_1 \otimes I_2.
\end{eqnarray} 
Here, $\mathbf{\hat{p}}_j$ denotes the canonical momentum operator of the $j$th particle,
$m_j$ the mass of the $j$th particle,
$\hat{\pmb{\sigma}}_j$ a vector that 
contains the three Pauli matrices of the $j$th particle,
and $\mathbf{r}_j$ the position vector of the $j$th particle.
In Eq.~\eqref{hamiltonian_tot},
$I_j$ denotes the $2$ by $2$ identity matrix
that spans the Hilbert space of the spin degrees of freedom
of the $j$th particle.
Defining the center-of-mass and relative coordinates $\mathbf{R}$ and $\mathbf{r}$,
$\mathbf{R}=(m_1\mathbf{r}_1+m_2\mathbf{r}_2)/(m_1+m_2)$ and $\mathbf{r}=\mathbf{r}_1-\mathbf{r}_2$,
Eq.~\eqref{hamiltonian_tot} can be rewritten as
\begin{eqnarray}
\label{hamiltonian_tot_com_rel}
\hat{H}_{\text{tot}}=\hat{H}_{\text{com}}^0+\hat{H}_{\text{rel}}^0+
\hat{V}_{\text{2b}}(\mathbf{r})I_1 \otimes I_2,
\end{eqnarray}
where
\begin{eqnarray}
\label{hamiltonian_com}
\hat{H}_{\text{com}}^0=
\frac{\mathbf{\hat{P}}^2}{2M}I_1 \otimes I_2+
\frac{\hbar k_{\text{so}}}{M}\mathbf{\hat{P}}\cdot
(\hat{\pmb{\sigma}}_1 \otimes I_2+I_1 \otimes \hat{\pmb{\sigma}}_2)
\end{eqnarray}
and
\begin{eqnarray}
\label{hamiltonian_rel}
\hat{H}_{\text{rel}}^0&=&
\frac{\mathbf{\hat{p}}^2}{2\mu}I_1 \otimes I_2
\nonumber \\
&+&
\frac{\hbar k_{\text{so}}}{\mu}\mathbf{\hat{p}}
\cdot\left(\frac{m_2\hat{\pmb{\sigma}}_1 \otimes I_2 -
m_1 I_1 \otimes \hat{\pmb{\sigma}}_2}{M}\right).
\end{eqnarray}
Here, 
$M$ and $\mu$ denote the center-of-mass and reduced masses,
$M=m_1+m_2$ and $\mu=m_1m_2/(m_1+m_2)$, 
and $\mathbf{\hat{P}}$ and $\mathbf{\hat{p}}$ 
are the center-of-mass 
and relative momentum operators,
$\mathbf{\hat{P}}=\mathbf{\hat{p}}_1+\mathbf{\hat{p}}_2$
and
$\mathbf{\hat{p}}=(m_2\mathbf{\hat{p}}_1-m_1\mathbf{\hat{p}}_2)/M$.
Since the system Hamiltonian $\hat{H}_{\text{tot}}$ commutes with 
$\mathbf{\hat{P}}$~\cite{comment_on_permute}, 
the center-of-mass momentum is conserved.
Throughout this paper,
we consider the situation where the expectation value of 
the center-of-mass momentum vanishes.
Integrating out the center-of-mass degrees of freedom,
the Hamiltonian $\hat{H}_{\text{tot}}$ reduces to 
\begin{eqnarray}
\label{Hamiltonian_tot_zero_com}
\hat{H}=\hat{H}_{\text{rel}}^0+
\hat{V}_{\text{2b}}(\mathbf{r})I_1 \otimes I_2.
\end{eqnarray} 

We start our discussion by considering 
the non-interacting Hamiltonian $\hat{H}_{\text{rel}}^0$.
Defining 
\begin{eqnarray}
\hat{\pmb{\Sigma}}=(m_2\hat{\pmb{\sigma}}_1 \otimes I_2 -
m_1 I_1 \otimes \hat{\pmb{\sigma}}_2)/M
\end{eqnarray}
and the relative helicity operator $\hat{h}_{\text{rel}}$,
\begin{eqnarray}
\label{hrel}
\hat{h}_{\text{rel}}=\frac{\hat{\mathbf{p}}\cdot\hat{\pmb{\Sigma}}}{|\langle\hat{\mathbf{p}}\cdot\hat{\pmb{\Sigma}}\rangle|},
\end{eqnarray}
Eq.~\eqref{hamiltonian_rel} can be rewritten as 
\begin{eqnarray}
\label{hamiltonian_rel_hrel}
\hat{H}_{\text{rel}}^0=
\frac{\hat{\mathbf{p}}^2}{2\mu}I_1 \otimes I_2+
\frac{\hbar k_{\text{so}}}{\mu}\hat{h}_{\text{rel}}|\langle\hat{\mathbf{p}}\cdot\hat{\pmb{\Sigma}}\rangle|.
\end{eqnarray}
The expectation value $|\langle\hat{\mathbf{p}} \cdot \hat{\mathbf{\Sigma}} \rangle |$
in the denominator of $\hat{h}_{\text{rel}}$,
which serves as a ``normalization factor'',
is evaluated with respect to the same state as the expectation value of 
$\hat{H}_{\text{rel}}^0$.
Since $\hat{h}_{\text{rel}}$ commutes with $\hat{H}_{\text{rel}}^0$,
the eigenstates and eigenvalues of $\hat{H}_{\text{rel}}^0$ can 
be labeled by the eigenvalues $h_{\text{rel}}$ of $\hat{h}_{\text{rel}}$,
where
$h_{\text{rel}}$ can take
the values $1$ 
and $-1$.

To determine the eigenenergies of $\hat{H}_{\text{rel}}^0$,
we consider a fixed relative momentum quantum number $\mathbf{p}$
and define $\mathbf{p}=\hbar\mathbf{k}$ and $k=|\mathbf{k}|$.
For a state with fixed $\mathbf{p}$ and $h_{\text{rel}}$,
we have
$|\langle\hat{\mathbf{p}}\cdot\hat{\pmb{\Sigma}}\rangle|=\hbar k|\langle\hat{\pmb{\Sigma}}\rangle|$~\cite{expectation_value_argument}.
Since $|\langle\hat{\pmb{\Sigma}}\rangle|$ can take the values
$1$ and $\eta$, 
where 
\begin{eqnarray}
\eta=(m_2-m_1)/M,
\end{eqnarray}
the eigenenergies of $\hat{H}_{\text{rel}}^0$
are
\begin{eqnarray}
\label{branch_A}
E_A=\frac{\hbar^2(\bar{k}+k_{\text{so}})^2}{2\mu}-E_r
\end{eqnarray}
and 
\begin{eqnarray}
\label{branch_B}
E_B=\frac{\hbar^2(\bar{k}+\eta k_{\text{so}})^2}{2\mu}-\eta^2E_r.
\end{eqnarray}
Here, $\bar{k}=h_{\text{rel}}k$ and $E_r$ is the recoil energy, $E_r=\hbar^2k_{\text{so}}^2/(2\mu)$.
The energies $E_A$ and $E_B$ are referred to as the energies of branch A and branch B,
respectively. 

\begin{figure}
\vspace*{0.1cm}
\hspace*{-0.4cm}
\includegraphics[width=0.4\textwidth]{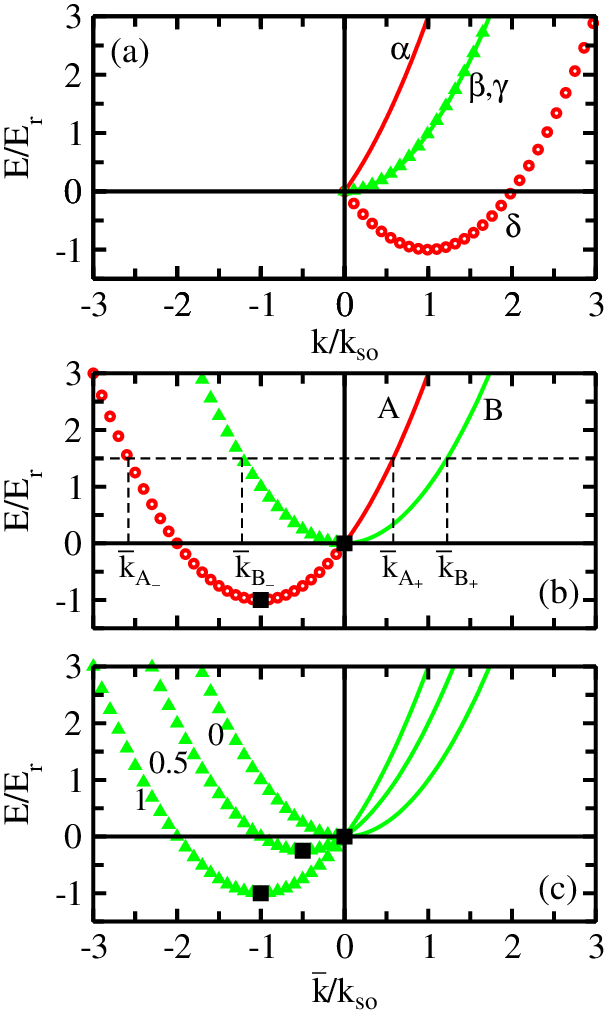}
\caption{
(Color online)
Dispersion relationships for $\hat{H}_{\text{rel}}^0$.
The energies $E_A$ (red solid lines and circles) and $E_B$ 
(green solid lines and triangles) are shown for $\eta=0$
as a function of $k$ 
in panel~(a) and as a function of $\bar{k}$ in panel (b);
the solid lines correspond to $h_{\text{rel}}=1$ and the symbols
correspond to
$h_{\text{rel}}=-1$.
For a fixed positive energy [see the horizontal dashed line in panel~(b)],
there exist four real solutions for $\bar{k}$,
denoted by $\bar{k}_{A_+}$, $\bar{k}_{A_-}$, $\bar{k}_{B_+}$, and 
$\bar{k}_{B_-}$
[see the vertical dashed lines and the discussion after Eq.~\eqref{mathcal_N}];
for $-E_r<E<0$, there exist only two real solutions for 
$\bar{k}$.
Panel (c) shows the energy $E_B$ for $\eta=0,1/2$ and $1$ (see the labels
in the figure).
In panels~(b) and (c), the filled squares show the 
minima $\bar{k}_A^{\text{min}}$ and $\bar{k}_B^{\text{min}}$.}
\label{fig1}
\end{figure}

Figure~\ref{fig1}(a) shows the eigenenergies 
of $\hat{H}_{\text{rel}}^0$ as a function of $k$ for equal masses (i.e., for $\eta=0$).
These dispersion curves are shown in many papers, including those
discussing two-body scattering~\cite{xiaoling,Gao}.
Since $k$ is by definition positive,
only the positive side of the horizontal axis exists.
The curves labeled by $\alpha$ and $\delta$ show $E_A$
and 
the curves labeled by $\beta$ and $\gamma$ show $E_B$; 
note that the curves labeled by $\beta$ and 
$\gamma$ are---for the equal-mass case shown---degenerate.
In Fig.~\ref{fig1}(a), 
the red and green solid lines show the energy of states
with
positive relative helicity ($h_{\text{rel}}=1$)
while the circles and triangles
show the energy of states
with
negative relative helicity ($h_{\text{rel}}=-1$).
Figure~\ref{fig1}(b) replots the eigenenergies as a function of
$\bar{k}$. Since $\bar{k}=h_{\text{rel}} k$,
the eigenenergies corresponding to states with negative
helicity (symbols) are, compared to Fig.~\ref{fig1}(a), 
``flipped'' to negative $\bar{k}$.
While there exist four states
for a fixed $k$ in Fig.~\ref{fig1}(a),
there exist two states
for a fixed $\bar{k}$ in Fig.~\ref{fig1}(b).
In this representation, branch A and branch B each correspond
to smooth parabola.
Figure~\ref{fig1}(c) illustrates the mass dependence of the energy of 
branch B.
For a finite mass imbalance, the minimum of branch B
is located at finite negative $\bar{k}$.
Specifically, as $\eta$ increases from 0 to 1,
the minimum $\bar{k}_B^{\text{min}}$ of $E_B$ moves  
from $0$ to $-k_{\text{so}}$.
For $\eta=1$ (infinite  mass imbalance), branch B is degenerate with
branch A.
Since branch $A$ is independent of $\eta$,
the minimum $\bar{k}_{A}^{\text{min}}$ of $E_A$ does not move
as $\eta$ changes.
The minima $\bar{k}^{\text{min}}_{A}$ and $\bar{k}^{\text{min}}_{B}$ are shown
by the filled black squares in Figs.~\ref{fig1}(b) and \ref{fig1}(c).

To obtain the eigenstates of $\hat{H}_{\text{rel}}^0$,
we write $\mathbf{k}$ in spherical coordinates,
$\mathbf{k}=k\mathbf{n}$ with $k$ the magnitude of $\mathbf{k}$ and
$\mathbf{n}$ the unit vector in the direction of $\mathbf{k}$,
$\mathbf{n}=\left(\sin\theta_{\mathbf{k}}\cos\phi_{\mathbf{k}},\sin\theta_{\mathbf{k}}\sin\phi_{\mathbf{k}},\cos\theta_{\mathbf{k}}\right)$,
and diagonalize the 4 by 4 Hamiltonian matrix in spin space.
The resulting eigenstates of the branches labeled by $\alpha$, $\beta$, $\gamma$, and $\delta$
in 
Fig.~\ref{fig1}(a) are
\begin{eqnarray}
\label{plane_wave_alpha}
|\alpha\rangle^{+}&=&\frac{e^{\imath\mathbf{k}\cdot\mathbf{r}}}{(2\pi)^{3/2}}
\Bigg(-\frac{1}{\sqrt{2}}e^{\imath\phi_\mathbf{k}}|\chi_{0,0}\rangle
-
\frac{\sin\theta_\mathbf{k}}{2}e^{2\imath\phi_\mathbf{k}}|\chi_{1,-1}\rangle
\nonumber \\
&-&
\frac{\cos\theta_\mathbf{k}}{\sqrt{2}}e^{\imath\phi_\mathbf{k}}|\chi_{1,0}\rangle+
\frac{\sin\theta_\mathbf{k}}{2}|\chi_{1,1}\rangle
\Bigg),
\end{eqnarray}
\begin{eqnarray}
\label{plane_wave_delta}
|\delta\rangle^{-}&=&\frac{e^{\imath\mathbf{k}\cdot\mathbf{r}}}{(2\pi)^{3/2}}
\Bigg(\frac{1}{\sqrt{2}}e^{\imath\phi_\mathbf{k}}|\chi_{0,0}\rangle-
\frac{\sin\theta_\mathbf{k}}{2}e^{2\imath\phi_\mathbf{k}}|\chi_{1,-1}\rangle
\nonumber \\
&-&
\frac{\cos\theta_\mathbf{k}}{\sqrt{2}}e^{\imath\phi_\mathbf{k}}|\chi_{1,0}\rangle+
\frac{\sin\theta_\mathbf{k}}{2}|\chi_{1,1}\rangle
\Bigg),
\end{eqnarray}
\begin{eqnarray}
\label{plane_wave_beta}
|\beta\rangle^{+}&=&\frac{e^{\imath\mathbf{k}\cdot\mathbf{r}}}{(2\pi)^{3/2}}
\Bigg[\sin^{2}\left(\frac{\theta_\mathbf{k}}{2}\right)e^{2\imath\phi_\mathbf{k}}|\chi_{1,-1}\rangle
\nonumber \\
&+&
\frac{\sin\theta_\mathbf{k}}{\sqrt{2}}e^{\imath\phi_\mathbf{k}}|\chi_{1,0}\rangle+
\cos^{2}\left(\frac{\theta_\mathbf{k}}{2}\right)|\chi_{1,1}\rangle
\Bigg],
\end{eqnarray}
and
\begin{eqnarray}
\label{plane_wave_gamma}
|\gamma\rangle^{-}&=&\frac{e^{\imath\mathbf{k}\cdot\mathbf{r}}}{(2\pi)^{3/2}}
\Bigg[\cos^{2}\left(\frac{\theta_\mathbf{k}}{2}\right)e^{2\imath\phi_\mathbf{k}}|\chi_{1,-1}\rangle
\nonumber \\
&-&
\frac{\sin\theta_\mathbf{k}}{\sqrt{2}}e^{\imath\phi_\mathbf{k}}|\chi_{1,0}\rangle+
\sin^{2}\left(\frac{\theta_\mathbf{k}}{2}\right)|\chi_{1,1}\rangle
\Bigg].
\end{eqnarray}
Here, the superscripts ``$+$'' and ``$-$'' indicate the relative helicity (``$+$'' corresponds
to $h_{\text{rel}}=1$ and ``$-$'' to
$h_{\text{rel}}=-1$).
Our goal is now to combine the states $|\alpha \rangle^+$
and $| \delta \rangle^-$ into a single state, the eigenstate
$|\bar{k},\theta_{\mathbf{k}},\phi_{\mathbf{k}}\rangle_{A}$ of branch A.
Inspection of Eqs.~(\ref{plane_wave_alpha}) and (\ref{plane_wave_delta})
shows that the spatial parts are not changing smoothly when the 
relative helicity  
changes from ``$+$'' to ``$-$'' (the spatial part associated with $|\chi_{0,0}\rangle$ changes sign).
Since the eigenstate $|\delta\rangle^{-}$
depends parametrically on $\mathbf{k}$,
the state given 
in Eq.~(\ref{plane_wave_delta})
remains an eigenstate if we change $\mathbf{k}$ 
to $-\mathbf{k}$.
Making this replacement and using the identities
$\theta_{-\mathbf{k}}=\pi-\theta_{\mathbf{k}}$ and $\phi_{-\mathbf{k}}=\pi+\phi_{\mathbf{k}}$,
Eq.~\eqref{plane_wave_delta} becomes
\begin{eqnarray}
\label{plane_wave_delta2}
|\delta\rangle^{-}&=&\frac{e^{-\imath\mathbf{k}\cdot\mathbf{r}}}{(2\pi)^{3/2}}
\Bigg(-\frac{1}{\sqrt{2}}e^{\imath\phi_\mathbf{k}}|\chi_{0,0}\rangle-
\frac{\sin\theta_\mathbf{k}}{2}e^{2\imath\phi_\mathbf{k}}|\chi_{1,-1}\rangle
\nonumber \\
&-&
\frac{\cos\theta_\mathbf{k}}{\sqrt{2}}e^{\imath\phi_\mathbf{k}}|\chi_{1,0}\rangle+
\frac{\sin\theta_\mathbf{k}}{2}|\chi_{1,1}\rangle
\Bigg)
.
\end{eqnarray}
Using $\bar{k}=h_{\text{rel}} k$,
Eqs.~\eqref{plane_wave_alpha} and~\eqref{plane_wave_delta2} can be combined 
to yield the eigenstate $|\bar{k},\theta_{\mathbf{k}},\phi_{\mathbf{k}}\rangle_{A}$
of branch A,
\begin{eqnarray}
\label{plane_wave_A}
|\bar{k},\theta_{\mathbf{k}},\phi_{\mathbf{k}}\rangle_{A}
&=&\frac{e^{\imath\bar{k}\mathbf{n}\cdot\mathbf{r}}}{(2\pi)^{3/2}}
\times
\nonumber \\
&&\Bigg(-\frac{1}{\sqrt{2}}e^{\imath\phi_\mathbf{k}}|\chi_{0,0}\rangle-
\frac{\sin\theta_\mathbf{k}}{2}e^{2\imath\phi_\mathbf{k}}|\chi_{1,-1}\rangle
\nonumber \\
&&-
\frac{\cos\theta_\mathbf{k}}{\sqrt{2}}e^{\imath\phi_\mathbf{k}}|\chi_{1,0}\rangle+
\frac{\sin\theta_\mathbf{k}}{2}|\chi_{1,1}\rangle
\Bigg).
\end{eqnarray}
For a fixed $\mathbf{n}$,
the branch A state changes smoothly 
from a state with ``$+$'' relative helicity to a state with ``$-$'' relative helicity
if the quantity $\bar{k}$
goes through zero.
The state has positive relative helicity if
$\bar{k}\mathbf{n}$ is parallel to $\langle\hat{\mathbf{\Sigma}}\rangle$
and negative relative helicity 
if $\bar{k}\mathbf{n}$ is anti-parallel to $\langle\hat{\mathbf{\Sigma}}\rangle$.
The eigenstate
$|\bar{k},\theta_{\mathbf{k}},\phi_{\mathbf{k}}\rangle_{B}$
of branch B with energy $E_B$ can be obtained by applying an analogous logic to
Eqs.~\eqref{plane_wave_beta} and~\eqref{plane_wave_gamma}.
We find
\begin{eqnarray}
\label{plane_wave_B}
|\bar{k},\theta_{\mathbf{k}},\phi_{\mathbf{k}}\rangle_{B}
&&=\frac{e^{\imath\bar{k}\mathbf{n}\cdot\mathbf{r}}}{(2\pi)^{3/2}} 
\Bigg[\sin^{2}\left(\frac{\theta_\mathbf{k}}{2}\right)e^{2\imath\phi_\mathbf{k}}|\chi_{1,-1}\rangle+
\nonumber \\
&&\frac{\sin\theta_\mathbf{k}}{\sqrt{2}}e^{\imath\phi_\mathbf{k}}|\chi_{1,0}\rangle
+
\cos^{2}\left(\frac{\theta_\mathbf{k}}{2}\right)|\chi_{1,1}\rangle
\Bigg].
\end{eqnarray}
The expectation 
value of $\hat{h}_{\text{rel}}$ 
with respect to $|\bar{k}, \theta_{\mathbf{k}},\phi_{\mathbf{k}}\rangle_{A}$
and $|\bar{k}, \theta_{\mathbf{k}},\phi_{\mathbf{k}}\rangle_{B}$
is
$\bar{k}/|\bar{k}|$,
which 
equals
$+1$ for $\bar{k}>0$
(positive helicity) 
and $-1$ 
for $\bar{k}<0$
(negative helicity).

We now 
construct, 
following the logic of Refs.~\cite{Gao,chris}, 
the
eigenstates of $\hat{H}_{\text{rel}}^0$
with good total angular momentum quantum number $J$
and corresponding projection quantum number $M_J$.
$\hat{H}_{\text{rel}}^0$ commutes with the square of the 
total angular momentum operator $\hat{\mathbf{J}}$
and its $z$-component $\hat{J}_z$,
where $\hat{\mathbf{J}}^2=(\hat{\mathbf{L}}+\hat{\mathbf{S}})^2$ and $\hat{J}_z=\hat{L}_z+\hat{S}_z$.
Here, $\hat{\mathbf{L}}$ is the relative 
orbital angular momentum operator of the two-particle system.
Using standard angular momentum 
algebra~\cite{zare},
the eigenstates of $\hat{\mathbf{J}}^2$ and $\hat{J}_z$ 
are constructed by taking linear combinations of states 
with different orbital angular momentum 
and spin angular momentum quantum numbers,
\begin{eqnarray}
\label{channel}
|J, M_J; L, S\rangle =\sum_{M_L, M_S} c_{M_L,M_S}^{J, M_J, L, S}|L, M_L; S, M_S\rangle,
\end{eqnarray}
where the $c_{M_L,M_S}^{J, M_J, L, S}$ denote Clebsch-Gordan coefficients
and 
$M_L$ and $M_S$
the projection quantum numbers corresponding
to the operators 
$\hat{L}_z$ and $\hat{S}_z$, 
respectively.
For each $(J,M_J)$ channel, 
we expand the eigenstates $\psi^{(J,M_J)}(\bar{k} \mathbf{r})$
of $\hat{H}_{\text{rel}}^0$ into components labeled by $(J,M_J;L,S)$, 
\begin{eqnarray}
\label{psi}
\psi^{(J,M_J)}(\bar{k} \mathbf{r})=\sum_{L,S}\frac{u_{L, S}^{(J,M_J)}(\bar{k}r)}{r}|J, M_J; L, S\rangle,
\end{eqnarray}
where the 
sums
over $L$ and $S$ 
go
over all 
quantum numbers allowed by angular momentum coupling and
where the ``weights'' $u^{(J,M_J)}_{L,S}(\bar{k} r)/r$ 
have the unit of inverse length and depend on the distance $r$. 
To obtain the weights for $\psi^{(J,M_J)}(\bar{k} \mathbf{r})$ 
(assuming that $J$ and $M_J$ are fixed),
we project the plane wave solutions, Eqs.~\eqref{plane_wave_A}-\eqref{plane_wave_B},
onto 
each $(J, M_J; L, S)$ component. 
This projection yields the following general structure for $\psi^{(J,M_J)}(\bar{k} \mathbf{r})$,
\begin{eqnarray}
\label{bessel_j}
\psi_{\text{reg}}^{(J,M_J)}(
\bar{k}
\mathbf{r})=\sum_{L,S}a_{L,S}^{J,M_J}\bar{k}j_{L}(\bar{k}r)|J, M_J; L, S\rangle,
\end{eqnarray}
where $j_{L}(\bar{k}r)$ is the spherical Bessel function of order $L$. 
Here, the coefficients $a_{L,S}^{J,M_J}$ are normalized
such that $\sum_{L,S} |a_{L,S}^{J,M_J}|^2=1$.
Equation~\eqref{bessel_j} constitutes the regular eigenstates of $\hat{H}_{\text{rel}}^0$.
To span the full Hilbert space,
a set of irregular solutions is needed in addition to the regular solutions.
Replacing the spherical Bessel functions $j_{L}(\bar{k}r)$ in Eq.~\eqref{bessel_j}
by spherical Neumann functions $n_L(\bar{k}r)$, 
we obtain the irregular solutions
\begin{eqnarray}
\label{bessel_n}
\psi_{\text{irr}}^{(J,M_J)}(
\bar{k}
\mathbf{r})=\sum_{L,S}a_{L,S}^{J,M_J}\bar{k}n_{L}(\bar{k}r)|J,M_J;L,S\rangle.
\end{eqnarray}
Looking ahead, we define 
\begin{eqnarray}
\label{psi_vector_reg}  
\underbar{$\psi$}_{\text{reg}}^{(J,M_J)}(\bar{k}r)=\left(a_{L_1,S_1}^{J,M_J}\bar{k}j_{L_1}(\bar{k}r), a_{L_2, S_2}^{J,M_J}\bar{k}j_{L_2}(\bar{k}r),...\right)^T
\end{eqnarray}
and
\begin{eqnarray}
\label{psi_vector_irr}  
\underbar{$\psi$}_{\text{irr}}^{(J,M_J)}(\bar{k}r)=\left(a_{L_1,S_1}^{J,M_J}\bar{k}n_{L_1}(\bar{k}r), a_{L_2, S_2}^{J,M_J}\bar{k}n_{L_2}(\bar{k}r),...\right)^T.
\end{eqnarray}

We now consider a non-vanishing short-range potential $\hat{V}_{\text{2b}}(\mathbf{r})$.
For a spherically symmetric two-body interaction potential $\hat{V}_{\text{2b}}(\mathbf{r})$,
the Hamiltonian $\hat{H}$, 
Eq.~\eqref{Hamiltonian_tot_zero_com},
commutes with $\hat{\mathbf{J}}^2$ and $\hat{J}_z$.
Thus we seek
scattering solutions for each $(J,M_J)$ channel.
The radial scattering wave function $\Psi^{(J,M_J)}(
r)$ in the $(J,M_J)$ 
channel for a fixed energy 
is written asymptotically, in the large $r$ limit, as
\begin{eqnarray}
\label{wave_function}
\Psi^{(J,M_J)}(r)\xrightarrow{r\to\infty}\mathcal{J}^{(J,M_J)}(r)-
\mathcal{N}^{(J,M_J)}(r)\mathcal{K}^{(J,M_J)},
\end{eqnarray}
where 
\begin{align}
\label{mathcal_J}
\mathcal{J}^{(J,M_J)}(r)=\nonumber \\
\Big[&N_{A_+}\underbar{$\psi$}_{\text{reg}_A}^{(J,M_J)}(\bar{k}_{A_+}r), 
N_{A_-}\underbar{$\psi$}_{\text{reg}_A}^{(J,M_J)}(\bar{k}_{A_-}r),\nonumber \\
&N_{B_+}\underbar{$\psi$}_{\text{reg}_B}^{(J,M_J)}(\bar{k}_{B_+}r), 
N_{B_-}\underbar{$\psi$}_{\text{reg}_B}^{(J,M_J)}(\bar{k}_{B_-}r)\Big]
\end{align}
and
\begin{align}
\label{mathcal_N}
\mathcal{N}^{(J,M_J)}(r)= \nonumber \\
\Big[&N_{A_+}\underbar{$\psi$}_{\text{irr}_A}^{(J,M_J)}(\bar{k}_{A+}r), 
-N_{A_-}\underbar{$\psi$}_{\text{irr}_A}^{(J,M_J)}(\bar{k}_{A-}r),\nonumber \\
&N_{B_+}\underbar{$\psi$}_{\text{irr}_B}^{(J,M_J)}(\bar{k}_{B+}r), 
-N_{B_-}\underbar{$\psi$}_{\text{irr}_B}^{(J,M_J)}(\bar{k}_{B-}r)\Big].
\end{align}
Here,
the subscripts $A$ and $B$ are added 
to the wave functions $\underbar{$\psi$}_{\text{reg}}^{(J,M_J)}(\bar{k}r)$
and 
$\underbar{$\psi$}_{\text{irr}}^{(J,M_J)}(\bar{k}r)$,
Eqs.~\eqref{psi_vector_reg} and~\eqref{psi_vector_irr},
to indicate the regular and irregular solutions of branch A and branch B,
respectively.
The subscripts $A_+$ and $A_-$ 
distinguish the two allowed $\bar{k}$ of branch A at fixed energy.
We use the convention
$\bar{k}_{A_+}>\bar{k}_A^{\text{min}}$ and $\bar{k}_{A_-}<\bar{k}_A^{\text{min}}$.
Similarly, the subscripts $B_+$ and  $B_-$ 
distinguish the two allowed $\bar{k}$ of branch B at fixed energy.
We use the convention
$\bar{k}_{B_+}>\bar{k}_B^{\text{min}}$ and $\bar{k}_{B_-}<\bar{k}_B^{\text{min}}$
[see Fig.~\ref{fig1}(b) for an illustration].
The $N_{A_+}$, $N_{A_-}$, $N_{B_+}$, and $N_{B_-}$ denote current 
normalization factors
that are chosen such that the
flux over a closed surface encircling the origin 
for a state with outgoing current is
normalized to $1$
[see Eqs.~(\ref{current_normalization}) and (\ref{current_normalization1}) below
and 
Ref.~\cite{note_on_normalization}].
The forms of $\mathcal{J}^{(J,M_J)}(r)$ and $\mathcal{N}^{(J,M_J)}(r)$ 
in Eqs.~(\ref{mathcal_J}) and (\ref{mathcal_N}) 
ensure that the scattering matrix $\mathcal{S}^{(J,M_J)}$ 
is related to the reaction matrix $\mathcal{K}^{(J,M_J)}$ in the ``usual way''
when the outgoing current boundary condition at large $r$ is matched (see the discussion below),
namely through
\begin{eqnarray}
\label{mathcal_S}
\mathcal{S}^{(J,M_J)}=(I+\imath\mathcal{K}^{(J,M_J)})(I-\imath\mathcal{K}^{(J,M_J)})^{-1}.
\end{eqnarray}
Here, $I$ denotes the identity matrix.
The reaction matrix $\mathcal{K}^{(J,M_J)}$ 
is determined by matching the 
radial
wave functions to the asymptotic 
form, Eq.~\eqref{wave_function}, 
at sufficiently large interparticle distances.
Specifically, 
if we know the large $r$ behavior of the wave function $\Psi^{(J,M_J)}(r)$ 
in the $|J,M_J;L,S\rangle$ basis for a short-range interaction potential 
$\hat{V}_{\text{2b}}(\mathbf{r})$,
then we can match its asymptotic large $r$ behavior to Eq.~\eqref{wave_function}.
In general,
$\Psi^{(J,M_J)}(r)$ can be obtained by propagating the logarithmic derivative matrix using an 
appropriate propagation scheme.
Importantly,
since Eqs.~(\ref{mathcal_J}) and (\ref{mathcal_N}) 
are continuous with respect to the arguments $\bar{k}_{A_+}, \bar{k}_{A_-}, \bar{k}_{B_+}$, and $\bar{k}_{B_-}$
and
since these arguments change smoothly 
when the
energy 
goes
from small positive to small negative values,
$\Psi^{(J,M_J)}(r)$ changes smoothly as the energy goes from positive to negative values. 

To relate the scattering matrix $\mathcal{S}^{(J,M_J)}$ and the reaction matrix $\mathcal{K}^{(J,M_J)}$,
we analyze the outgoing currents at large distances.
The relative current $\mathbf{j}(\mathbf{r})$ for 
state $\psi^{(J,M_J)}(\bar{k}\mathbf{r})$ is defined through
\begin{eqnarray}
\label{current}
\mathbf{j}(\mathbf{r})=\frac{1}{2\mu}
\Bigg\{
(\psi^{(J,M_J)})^{\dagger}
\mathbf{p}_{k}
\psi^{(J,M_J)} 
+ 
\nonumber \\
\left[
(\psi^{(J,M_J)})^{\dagger}
\mathbf{p}_{k}
\psi^{(J,M_J)}
\right]^*\Bigg\},
\end{eqnarray}
where $\mathbf{p}_k$ is the relative kinetic momentum operator,
which is related to the relative canonical momentum operator $\mathbf{p}$ by 
$\mathbf{p}_k=\mu[\mathbf{r}, H_{\text{rel}}^0]/(\imath\hbar)$.
We find 
\begin{eqnarray}
\label{pk}
\mathbf{p}_k=\mathbf{p} 
I_1 \otimes I_2
+\hbar k_{\text{so}} \hat{\pmb{\Sigma}}. 
\end{eqnarray}
Recall, $\psi^{(J,M_J)}$ is a four-component spinor and each of the
three components of $\mathbf{p}_k$ 
is a $4$ 
by
$4$ matrix. Doing the multiplications
in Eq.~(\ref{current}), we can check that $\mathbf{j}(\mathbf{r})$ 
is---as it should be---a three-component vector.
To obtain scattering wave functions 
that correspond to
outgoing current at large $r$,
we consider the following two linear combinations of 
Eqs.~\eqref{bessel_j} and~\eqref{bessel_n},
\begin{eqnarray}
\label{outgoing_phase}
\psi^{(J,M_J)+}(\bar{k}\mathbf{r})=\imath\psi_{\text{reg}}^{(J,M_J)}(\bar{k}\mathbf{r})-\psi_{\text{irr}}^{(J,M_J)}(\bar{k}\mathbf{r})
\end{eqnarray}
and
\begin{eqnarray}
\label{incoming_phase}
\psi^{(J,M_J)-}(\bar{k}\mathbf{r})=-\imath\psi_{\text{reg}}^{(J,M_J)}(\bar{k}\mathbf{r})-\psi_{\text{irr}}^{(J,M_J)}(\bar{k}\mathbf{r}).
\end{eqnarray}
Here, $\psi^{(J,M_J)+}(\bar{k}\mathbf{r})$ contains the ``phase factor'' $\text{e}^{\imath\bar{k}r}$ 
while $\psi^{(J,M_J)-}(\bar{k}\mathbf{r})$ contains the ``phase factor'' $\text{e}^{-\imath\bar{k}r}$.
The currents for $\psi^{(J,M_J)+}(\bar{k}\mathbf{r})$ and  $\psi^{(J,M_J)-}(\bar{k}\mathbf{r})$ are
\begin{eqnarray}
\label{current_outgoing_phase1}
\mathbf{j}^+(\mathbf{r})=\frac{\hbar}{4\pi\mu r^2}(\bar{k}-\bar{k}^{\text{min}})\hat{\mathbf{r}}
\end{eqnarray}
and
\begin{eqnarray}
\label{current_incoming_phase2}
\mathbf{j}^-(\mathbf{r})=-\frac{\hbar}{4\pi\mu r^2}(\bar{k}-\bar{k}^{\text{min}})\hat{\mathbf{r}},
\end{eqnarray}
respectively,
where $\bar{k}^{\text{min}}$ is the $\bar{k}$ value
at which the energy is minimal and $\hat{\mathbf{r}}$ is the unit vector in the $\mathbf{r}$ direction.
Importantly, the states with 
outgoing current
are $\psi^{(J,M_J)+}(\bar{k}\mathbf{r})$ 
for $\bar{k}>\bar{k}^{\text{min}}$
and
$\psi^{(J,M_J)-}(\bar{k}\mathbf{r})$ 
for $\bar{k}<\bar{k}^{\text{min}}$.
The current normalization factors
are determined by
enforcing current conservation, i.e., by enforcing
\begin{eqnarray}
\label{current_normalization}
|N|^2\oint\mathbf{j}^+(\mathbf{r})\cdot\text{d}\mathbf{S}=1 \text{ for }\ \bar{k}>\bar{k}^{\text{min}}
\end{eqnarray}
and
\begin{eqnarray}
\label{current_normalization1}
|N|^2\oint\mathbf{j}^-(\mathbf{r})\cdot\text{d}\mathbf{S}=1 \text{ for }\ \bar{k}<\bar{k}^{\text{min}}.
\end{eqnarray}
Table~\ref{table:current} shows the current and 
the corresponding current normalization factor for the states
$\psi_A^{(J,M_J)+}$, $\psi_A^{(J,M_J)-}$, $\psi_B^{(J,M_J)+}$, and $\psi_B^{(J,M_J)-}$.
\begin{table*}
\begin{center}
\begin{tabular}{c|c|c}
state
& $\mathbf{j}(\mathbf{r})$ & $N$\\
\hline
$\psi_{A}^{(J,M_J)+}(\bar{k}_{A_+}\mathbf{r})$ & $+\frac{\hbar}{4\pi\mu r^2}(\bar{k}_{A_+}-\bar{k}_A^{\text{min}})\hat{\mathbf{r}}$, out
&\multirow{2}{*}{$\sqrt{{\mu}/[{\hbar(\bar{k}_{A_+}-\bar{k}_A^{\text{min}})}]}$} \\
$\psi_{A}^{(J,M_J)-}(\bar{k}_{A_+}\mathbf{r})$ & $-\frac{\hbar}{4\pi\mu r^2}(\bar{k}_{A_+}-\bar{k}_A^{\text{min}})\hat{\mathbf{r}}$, in \\
\hline
$\psi_{A}^{(J,M_J)+}(\bar{k}_{A_-}\mathbf{r})$ & $+\frac{\hbar}{4\pi\mu r^2}(\bar{k}_{A_-}-\bar{k}_A^{\text{min}})\hat{\mathbf{r}}$, in 
&\multirow{2}{*}{$\sqrt{{\mu}{/[-\hbar(\bar{k}_{A_-}-\bar{k}_A^{\text{min}})]}}$} \\
$\psi_{A}^{(J,M_J)-}(\bar{k}_{A_-}\mathbf{r})$ & $-\frac{\hbar}{4\pi\mu r^2}(\bar{k}_{A_-}-\bar{k}_A^{\text{min}})\hat{\mathbf{r}}$, out \\
\hline
\hline
$\psi_{B}^{(J,M_J)+}(\bar{k}_{B_+}\mathbf{r})$ & $+\frac{\hbar}{4\pi\mu r^2}(\bar{k}_{B_+}-\bar{k}_B^{\text{min}})\hat{\mathbf{r}}$, out 
&\multirow{2}{*}{$\sqrt{{\mu}{/[\hbar(\bar{k}_{B_+}-\bar{k}_B^{\text{min}})]}}$} \\
$\psi_{B}^{(J,M_J)-}(\bar{k}_{B_+}\mathbf{r})$ & $-\frac{\hbar}{4\pi\mu r^2}(\bar{k}_{B_+}-\bar{k}_B^{\text{min}})\hat{\mathbf{r}}$, in\\
\hline
$\psi_{B}^{(J,M_J)+}(\bar{k}_{B_-}\mathbf{r})$ & $+\frac{\hbar}{4\pi\mu r^2}(\bar{k}_{B_-}-\bar{k}_B^{\text{min}})\hat{\mathbf{r}}$, in
&\multirow{2}{*}{$\sqrt{{\mu}{/[-\hbar(\bar{k}_{B_-}-\bar{k}_B^{\text{min}})]}}$} \\
$\psi_{B}^{(J,M_J)-}(\bar{k}_{B_-}\mathbf{r})$ & $-\frac{\hbar}{4\pi\mu r^2}(\bar{k}_{B_-}-\bar{k}_B^{\text{min}})\hat{\mathbf{r}}$, out \\
\end{tabular}
\caption{Columns 2 and 3 
report, respectively, the current and the corresponding current normalization 
factor
for the states listed in column 1.
The labels ``out'' and ``in'' in column 2 
indicate that the current points in the  $+\hat{\mathbf{r}}$ 
and $-\hat{\mathbf{r}}$ direction, respectively.
From top to bottom, the entries in column 3 correspond
to $N_{A_+}$, $N_{A_-}$, $N_{B_+}$,
 and $N_{B_-}$.
}
\label{table:current}
\end{center}
\end{table*}
For each branch A state, $\bar{k}$ takes the values $\bar{k}_{A_+}$ and $\bar{k}_{A_-}$;
similarly,
for each branch B state, $\bar{k}$ takes the values $\bar{k}_{B_+}$ and $\bar{k}_{B_-}$.
Since we have 
$\bar{k}_{A_+}>\bar{k}_A^{\text{min}}$, $\bar{k}_{A_-}<\bar{k}_A^{\text{min}}$, 
$\bar{k}_{B_+}>\bar{k}_B^{\text{min}}$, and $\bar{k}_{B_-}<\bar{k}_B^{\text{min}}$,
the current points in the
$+\hat{\mathbf{r}}$ direction for the states 
$\psi_{A}^{(J,M_J)+}(\bar{k}_{A_+}\mathbf{r})$,
$\psi_{A}^{(J,M_J)-}(\bar{k}_{A_-}\mathbf{r})$,
$\psi_{B}^{(J,M_J)+}(\bar{k}_{B_+}\mathbf{r})$,
and
$\psi_{B}^{(J,M_J)-}(\bar{k}_{B_-}\mathbf{r})$;
for the other four states, the current points in the
$-\hat{\mathbf{r}}$ direction.

To obtain the 
scattering matrix $\mathcal{S}^{(J,M_J)}$,
we rewrite Eq.~\eqref{wave_function} in terms of 
the states listed in the first column of Table~\ref{table:current}.
Grouping the states with outgoing current together 
and those with incoming current together,
we find
\begin{widetext}
\begin{align}
\label{wave_function_S}
&\Psi^{(J,M_J)}(\bar{k}r)\xrightarrow{r\to\infty}\nonumber \\
&\frac{1}{2\imath}\Big\{\Big[N_{A_+}\underbar{$\psi$}_{A}^{(J,M_J)+}(\bar{k}_{A_+}r), 
-N_{A_-}\underbar{$\psi$}_{A}^{(J,M_J)-}(\bar{k}_{A_-}r), 
N_{B_+}\underbar{$\psi$}_{B}^{(J,M_J)+}(\bar{k}_{B_+}r), 
-N_{B_-}\underbar{$\psi$}_{B}^{(J,M_J)-}(\bar{k}_{B_-}r)\Big]\mathcal{S}^{(J,M_J)}- \nonumber \\
&\Big[N_{A_+}\underbar{$\psi$}_{A}^{(J,M_J)-}(\bar{k}_{A+}r), 
-N_{A_-}\underbar{$\psi$}_{A}^{(J,M_J)+}(\bar{k}_{A-}r),
N_{B_+}\underbar{$\psi$}_{B}^{(J,M_J)-}(\bar{k}_{B+}r), 
-N_{B_-}\underbar{$\psi$}_{B}^{(J,M_J)+}(\bar{k}_{B-}r)\Big]\Big\} 
(I-\imath\mathcal{K}^{(J,M_J)}).
\end{align}
\end{widetext}

Important observables are the partial cross sections
$\sigma_{jl}^{(J,M_J)}$ that characterize the 
scattering from state $j$ to state $l$.
Throughout this paper,
state $j$ ($j=1-4$) corresponds to the state
in the $j$th column of $\mathcal{J}^{(J,M_J)}(r)$ and $\mathcal{N}^{(J,M_J)}(r)$.
This means that states 1-4 have the
arguments 
$\bar{k}_1=\bar{k}_{A_+}$, 
$\bar{k}_2=\bar{k}_{A_-}$, 
$\bar{k}_3=\bar{k}_{B_+}$, and 
$\bar{k}_4=\bar{k}_{B_-}$.
The partial cross sections~\cite{partial_cross_section_argument}
are related to the elements of the $\mathcal{S}^{(J,M_J)}$ matrix by
\begin{eqnarray}
\label{partial_cross_section}
\sigma_{jl}^{(J,M_J)}=\frac{2\pi}{\bar{k}_{j}^2}|\mathcal{S}_{lj}^{(J,M_J)}-\delta_{jl}|^2,
\end{eqnarray}
where $\delta_{jl}$ is the Kronecker delta function.
Since the lowest scattering threshold occurs at $E=-E_r$,
two-body bound states may exist for $E<-E_r$.
The bound state energies can be obtained by analyzing the poles of the $\mathcal{S}^{(J,M_J)}$ matrix.
Explicit calculations for the 
$(J,M_J)=(0,0)$ and $(1,M_J)$ channels are detailed in the next two sections.

\section{$(J,M_J)=(0,0)$ channel}
\label{sec3}
This section considers 
the $(J,M_J)=(0,0)$ channel.
Angular momentum algebra yields that 
the components $(J,M_J;L,S)=(0,0;0,0)$ and $(0,0;1,1)$ contribute.
Since the $(J,M_J;L,S)=(0,0;0,0)$ component
contains the spherical harmonic $Y_{0,0}$,
multiplied by the spin singlet, and
since the $(J,M_J;L,S)=(0,0;1,1)$ 
component 
contains the 
spherical harmonics $Y_{1,M_L}$ ($M_L=-1,0$, and $1$),
multiplied by spin triplets, 
both components are 
anti-symmetric 
under the exchange of the two particles.
Using the projection procedure outlined in the previous section,
we find that branch $A$ has
non-zero components in the $(J,M_J)=(0,0)$ channel
and that the last two columns in Eqs.~\eqref{mathcal_J} and~\eqref{mathcal_N}
are zero.
Thus, throughout this section, we drop the last two columns, 
i.e., we work with 
${\cal{K}}^{(0,0)}$ and ${\cal{S}}^{(0,0)}$ matrices of size 2 by 2 
that describe the physics of branch A.
Since  the dispersion relationship of branch $A$
is independent of the mass ratio, our results in this section
are independent of the mass ratio. 
Thus,
since
both contributing components are
anti-symmetric under the exchange of the two particles,
the scattering solutions obtained in this section
apply to either two identical fermions or to two distinguishable particles
with equal or unequal masses.
We have
\begin{eqnarray}
\label{mathcal_J_00}
\mathcal{J}^{(0,0)}(r)= \nonumber \\
\begin{bmatrix}
\frac{1}{\sqrt{2}}N_{A_+}\bar{k}_{A_+}j_0(\bar{k}_{A_+}r)
&\frac{1}{\sqrt{2}}N_{A_-}\bar{k}_{A_-}j_0(\bar{k}_{A_-}r)\\
-\frac{\imath}{\sqrt{2}}N_{A_+}\bar{k}_{A_+}j_1(\bar{k}_{A_+}r)
&-\frac{\imath}{\sqrt{2}}N_{A_-}\bar{k}_{A_-}j_1(\bar{k}_{A_-}r) 
\end{bmatrix},
\end{eqnarray}
\begin{eqnarray}
\label{mathcal_N_00}
\mathcal{N}^{(0,0)}(r)= \nonumber \\
\begin{bmatrix}
\frac{1}{\sqrt{2}}N_{A_+}\bar{k}_{A_+}n_0(\bar{k}_{A_+}r)
&-\frac{1}{\sqrt{2}}N_{A_-}\bar{k}_{A_-}n_0(\bar{k}_{A_-}r)\\
-\frac{\imath}{\sqrt{2}}N_{A_+}\bar{k}_{A_+}n_1(\bar{k}_{A_+}r)
&\frac{\imath}{\sqrt{2}}N_{A_-}\bar{k}_{A_-}n_1(\bar{k}_{A_-}r) 
\end{bmatrix},
\end{eqnarray}
and
\begin{eqnarray}
\label{wave_function_00}
\Psi^{(0,0)}(r)=\mathcal{J}^{(0,0)}(r)-\mathcal{N}^{(0,0)}(r)\mathcal{K}^{(0,0)}.
\end{eqnarray}
For positive energy, 
Eqs.~(\ref{mathcal_J_00}) and (\ref{mathcal_N_00}) 
agree
with Eqs.~(18) and (19) in Ref.~\cite{Gao}.

We now determine analytical results for
the two-body zero-range pseudo-potential $\hat{V}_{\text{ps}}(\mathbf{r})$,
\begin{eqnarray}
\label{V_ps} 
\hat{V}_{\text{ps}}(\mathbf{r})=\sum_{S}\frac{2\pi\hbar^2a_s}{\mu}
\delta^{(3)}
(\mathbf{r})\frac{\partial}{\partial r}r|J,M_J;0,S\rangle\langle J,M_J;0,S|,
\end{eqnarray}
where $a_s$ is the two-body free-space $s$-wave scattering length.
This model interaction has been used previously in 
Refs.~\cite{Gao,chris}.
The use of the zero-range potential limits the applicability of 
the results 
[see, e.g., Eqs.~\eqref{cross_section_00_1}-\eqref{cross_section_00_2} below] 
to situations where the effective range $r_0$ can be neglected,
i.e., situations where 
$r_0\ll \text{min}(1/k_{\text{so}},1/|\bar{k}_{A_+}|,1/|\bar{k}_{A_-}|, |a_s|)$,
and where higher-partial wave free-space scattering phase shifts 
can be neglected. The reason for this is
that Eq.~(\ref{V_ps}) depends on the free-space
$s$-wave scattering length but
not on the generalized 
$p$-wave
and higher partial wave scattering lengths.
If these conditions are fulfilled,
we expect that the results derived below within 
the zero-range $s$-wave pseudo-potential
approximation
reproduce observables for realistic van der Waals potentials 
in the presence of spin-orbit coupling quite accurately.
It is left to a future publication to quantify the agreement 
(or disagreement) between the zero-range and finite-range treatments. 

We stress that the zero-range pseudo-potential is used
differently here than in Ref.~\cite{xiaoling}.
As we discuss in more 
detail
below,
Ref.~\cite{xiaoling} did not start with a zero-range pseudo-potential
but instead derived an effective pseudo-potential description with effective coupling
constants that account for the modification
of the free-space coupling constants by the 
single-particle spin-orbit coupling terms.

The $\mathcal{K}^{(0,0)}$ matrix is determined by matching the $s$-
and $p$-wave boundary conditions,
i.e., 
the $s$-wave component of $\Psi^{(0,0)}(r)$ 
is forced to be proportional to
$1/r-1/a_s$ in the $r\to 0$ limit
and the $p$-wave component 
of $\Psi^{(0,0)}(r)$ 
is forced to have a vanishing
$1/r^2$ term in the $r\to 0$ limit~\cite{Gao, pengzhang, pengzhang1, zhenhua}.
The 
resulting
$\mathcal{K}^{(0,0)}$ matrix reads
\begin{eqnarray}
\label{mathcal_K_00}
\mathcal{K}^{(0,0)}=\frac{-a_s}{\bar{k}_{A_+}-\bar{k}_{A_-}}
\begin{bmatrix}
\bar{k}_{A_+}^2 &\bar{k}_{A_+}\bar{k}_{A_-}\\
\bar{k}_{A_+}\bar{k}_{A_-} &\bar{k}_{A_-}^2\\
\end{bmatrix}
.
\end{eqnarray}
The $\mathcal{S}^{(0,0)}$ matrix is obtained from the $\mathcal{K}^{(0,0)}$ matrix
using Eq.~\eqref{mathcal_S}.
The partial cross sections,
in turn,
are obtained using Eq.~\eqref{partial_cross_section}. 
We find 
\begin{eqnarray}
\label{cross_section_00_1}
\sigma_{11}^{(0,0)}=\sigma_{21}^{(0,0)}=
\frac{8\pi a_s^2 \bar{k}_{A_+}^2}{(\bar{k}_{A_+}-\bar{k}_{A_-})^2+a_s^2(\bar{k}_{A_+}^2+\bar{k}_{A_-}^2)^2}
\end{eqnarray}
and
\begin{eqnarray}
\label{cross_section_00_2}
\sigma_{12}^{(0,0)}=\sigma_{22}^{(0,0)}=
\left(\frac{\bar{k}_{A_-}}{\bar{k}_{A_+}}\right)^2 
\sigma_{11}^{(0,0)}.
\end{eqnarray}
Recall, 
$\bar{k}_{A_+}$ and $\bar{k}_{A_-}$ depend on the relative scattering energy $E$.
Thus, the partial cross sections given in Eqs.~\eqref{cross_section_00_1}-\eqref{cross_section_00_2}
contain the full energy-dependence. 
We stress that
Eqs.~\eqref{mathcal_K_00}-\eqref{cross_section_00_2} 
apply to all 
energies.
For positive energy,
we have $\bar{k}_{A_+}> 0$ and $\bar{k}_{A_-}<0$.
Our results for $E\ge 0$ agree with those presented in Ref.~\cite{Gao}.
For negative energy ($E > -E_r$),
we have
$\bar{k}_{A}^{\text{min}}< \bar{k}_{A_+}<0$ and $\bar{k}_{A_-}<\bar{k}_{A}^{\text{min}}$.
At the scattering threshold $E=-E_r$, $\bar{k}_{A_+}$ is equal to $\bar{k}_{A_-}$. 
Using this,
it can be seen readily from Eqs.~(\ref{cross_section_00_1}) 
and (\ref{cross_section_00_2}) that
all four partial cross sections approach the same value as
$E \rightarrow -E_r$.

\begin{figure}
\vspace*{0.6cm}
\hspace*{0cm}
\includegraphics[width=0.45\textwidth]{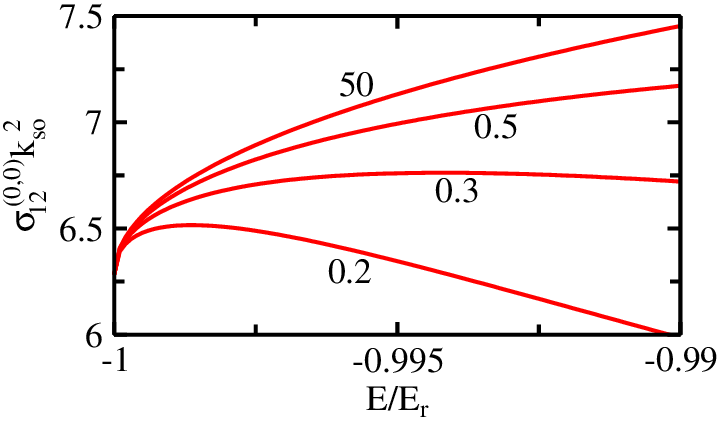}
\vspace*{0.4cm}
\caption{
(Color online)
Illustration of the threshold behavior 
for the $(J,M_J)=(0,0)$ channel. 
The solid lines show the partial cross 
sections $\sigma_{12}^{(0,0)}=\sigma_{22}^{(0,0)}$, 
in units of $1/k_{\text{so}}^2$,
as a function of the energy $E$ in the vicinity of $E=-E_r$ 
for four different $|a_sk_{\text{so}}|$,
$|a_sk_{\text{so}}|=0.2,0.3,0.5$ and $50$ (see labels in the figure).
The partial cross section 
$\sigma_{12}^{(0,0)}$
is equal to 
$2\pi /k_{\text{so}}^2$ 
at $E= -E_r$
for all
$|a_sk_{\text{so}}|$.
}
\label{fig2}
\end{figure}

To gain more insight,
we Taylor expand Eqs.~\eqref{cross_section_00_1} and~\eqref{cross_section_00_2} 
around the scattering threshold $E=-E_r$.
Near $E=-E_r$ (here, we exclude $a_s=0$),
we find that
the threshold behavior for all partial cross sections
is independent of $a_s$.
Table~\ref{table:threshold_00} shows the first three terms of the 
power series in terms of the small parameter 
$x=(E/E_r+1)^{1/2}$.
The partial cross sections 
$\sigma_{jl}^{(0,0)}$ 
approach the constant $2\pi/k_{\text{so}}^2$
for all finite $a_s$. 
This threshold behavior is distinctly different from 
the behavior near $E=0$ (see Ref.~\cite{Gao}) and
from the typical $s$-wave threshold law 
($\sigma\to 8\pi a_{s}^2$ for two identical bosons).
Figure~\ref{fig2} illustrates the threshold behavior near $E=-E_r$ for various
$|a_sk_{\text{so}}|$ combinations. 
Since the first-order correction is also independent of $a_s$ (see Table~\ref{table:threshold_00}),
the variation of $\sigma_{jl}^{(0,0)}$ near $E=-E_r$ is the same for all $a_s$.
\begin{table}
\begin{center}
\begin{tabular}{c|c}
&$E\to -E_r$ $(a_s \ne 0)$\\
\hline
$\sigma_{11}^{(0,0)}k_{\text{so}}^2=\sigma_{21}^{(0,0)}k_{\text{so}}^2$
&
$2\pi-4\pi x
-2\pi
\left[(a_sk_{\text{so}})^{-2}+1\right]
x^2 $ \\
$\sigma_{12}^{(0,0)}k_{\text{so}}^2=\sigma_{22}^{(0,0)}k_{\text{so}}^2$
&
$2\pi+4\pi x
-2\pi\left[ (a_sk_{\text{so}})^{-2}+1\right] x^2$ \\
\end{tabular}
\caption{
New threshold behavior
for two particles in the $(J,M_J)=(0,0)$ channel
near $E=-E_r$.
The partial cross sections
are independent of $a_s$ and approach, in units of $1/k_{\text{so}}^2$, 
the constant $2\pi$
at $E=-E_r$.
The 
first-order
correction is
also independent of $a_s$
for all $\sigma_{jl}^{(0,0)}$.
Here, we introduced the abbreviation 
$x=\sqrt{E/E_r+1}$.
}
\label{table:threshold_00}
\end{center}
\end{table}
We refer to this behavior as universal as it is fully determined by
the single-particle quantity $k_{\text{so}}$. 
The energy range over which the first-order correction provides a good description depends on $a_s$.
For large $|a_sk_{\text{so}}|$,
the second-order correction is small.
For small $|a_sk_{\text{so}}|$,
in contrast,
the second-order correction is significant,
leading to a ``turn-around'' of $\sigma_{jl}^{(0,0)}$
(see, e.g., the lowest curve in Fig.~\ref{fig2} for $|a_s k_{\text{so}}|=0.2$).
Since the isotropic three-dimensional spin-orbit coupling
scenario has not yet been realized experimentally in cold atom systems, we use
values from one-dimensional experimental realizations~\cite{spielman}
to get a feeling for the scattering length values covered
in Fig.~\ref{fig2}. 
Using 
$k_{\text{so}}=\sqrt{2}\pi/\lambda$ 
and 
$\lambda \approx 804nm$, 
the values of
$a_s k_{\text{so}}=0.2$ and $50$ correspond to 
$a_s \approx 680a_0$ and $170,000a_0$,
where $a_0$ denotes the 
Bohr
radius;
such scattering length values can be realized by utilizing Feshbach
resonance techniques~\cite{chin}. 
We also note that proposals for adjusting the spin-orbit
coupling strength have been put forward~\cite{tune_soc, tune_soc2, tune_soc3}.

The asymptotic 
large $r$
basis chosen in our work and in Ref.~\cite{xiaoling}
differ.
Reference~\cite{xiaoling} used a rotated basis, which 
is related to our asymptotic basis through the application of the
matrix $U$
[i.e., the matrices $\mathcal{J}^{(0,0)}(r)U$ and $\mathcal{N}^{(0,0)}(r)U$
are used instead of
$\mathcal{J}^{(0,0)}(r)$ and $\mathcal{N}^{(0,0)}(r)$
as in our work], where $U$ is
chosen such that
$U^{-1} \mathcal{K}^{(0,0)} U$
and $U^{-1} \mathcal{S}^{(0,0)} U$ are diagonal.
The incoming states of Ref.~\cite{xiaoling}
cannot be labeled by a fixed momentum and instead are 
linear combinations of states with different momenta.
Reference~\cite{zhenhua}, which also discussed the relationship
between the different asymptotic basis, referred to these 
states as ``standing waves''.
Diagonalizing $\mathcal{K}^{(0,0)}$,
one finds that
one of the eigenvalues of $\mathcal{K}^{(0,0)}$ 
is 0 (this corresponds to a vanishing
``effective''
$p$-wave phase shift $\delta_{\text{eff}}^p$, 
i.e., $\delta_{\text{eff}}^p=0$) and that the 
other 
eigenvalue
is
given 
by the ``effective''
$s$-wave phase shift
$\delta_{\text{eff}}^s$,
\begin{eqnarray}
\label{phase_shift}
\text{tan}\delta_{\text{eff}}^s=-a_sk_{\text{so}}
\left[\left(\frac{E}{E_r}+1\right)^{1/2}+\left(\frac{E}{E_r}+1\right)^{-1/2}\right]  .
\end{eqnarray}
Expression~(\ref{phase_shift}) agrees with Eq.~(44)
from Ref.~\cite{xiaoling}.
The fact that one of the eigenvalues is zero is a consequence of
the fact that our interaction potential $\hat{V}_{\text{ps}}$ 
does not account for
a
finite free-space $p$-wave phase 
shift.
While Ref.~\cite{zhenhua} interpreted the
phase shifts $\delta_{\text{eff}}^s$ and $\delta_{\text{eff}}^p$
as the equivalent of the usual free-space 
$s$- and $p$-wave phase shifts, we prefer to think of these
phase shifts as effective phase shifts that are obtained
by switching to the standing wave picture. The ``true'' phase
shifts, defined in analogy to the standard
partial wave decomposition, are those obtained
by writing the elements of the
matrix given in Eq.~(\ref{mathcal_K_00})
as $\tan \delta_{jl}$.

In the ``high energy'' regime ($E\gg -E_r$),
the first term in Eq.~\eqref{phase_shift}
dominates and
the tangent of the phase shift is approximately proportional to
the density of states of a three-dimensional system.
Near the scattering threshold ($E\to -E_r$),
the second term in Eq.~\eqref{phase_shift} dominates and
the tangent of the 
phase shift is approximately proportional to
the density of states of a one-dimensional system.
This effective dimensionality reduction is responsible
for the emergence of the new
universal threshold law (see Table~\ref{table:threshold_00}).
Equivalently, for $E\to -E_r$,
the eigenvalue
given in Eq.~(\ref{phase_shift})
diverges 
and both the $\mathcal{S}^{(0,0)}$ matrix
and the partial cross sections 
are independent of $a_s$.
As the above discussion shows, the analysis of the effective
phase shifts, obtained by switching to the ``standing wave'' basis,
provides a simple intuitive understanding of the modified
threshold law discussed in the context
of Table~\ref{table:threshold_00}.
Moreover, the standing wave basis does allow for the construction
of effective low-energy interactions that may be easier to deal
with in multi-body theories than the original 
pseudo-potential~\cite{xiaoling,zhenhua}.

To determine the bound state energies, 
we calculate the poles of the $\mathcal{S}^{(0,0)}$ matrix.
The bound state energy $E_{\text{bound}}^{(0,0)}$ is determined by 
the roots of the equation
\begin{eqnarray}
\label{equation_bound_energy_00}
\tan 
\delta_{\text{eff}}^s
=-\imath.
\end{eqnarray}
In agreement with
Refs.~\cite{bound_shenoy,xiaoling,pengzhang,zhenhua},
we find that
there exists a bound state for all $a_s$.
The bound state energy reads
\begin{eqnarray}
\label{bound_energy_00}
E_{\text{bound}}^{(0,0)}=-\frac{\hbar^2k_*^2}{2\mu}-E_r,
\end{eqnarray}
where
\begin{eqnarray}
\label{bound_energy_k*_00}
k_*=\frac{1+\text{sign}(a_s)\sqrt{1+4a_s^2k_{\text{so}}^2}}{2a_s}.
\end{eqnarray}
For $k_{\text{so}}=0$, Eqs.~(\ref{bound_energy_00}) and
(\ref{bound_energy_k*_00}) reduce to the binding energy
$-\hbar^2/(2 \mu a_s^2)$ for pure $s$-wave contact interactions with
positive $a_s$.

\section{$(J,M_J)=(1,M_J)$ channels}
\label{sec4}
This section considers two particles in the $(J,M_J)=(1,M_J)$ channels,
where $M_J$ takes the values 
$-1$, $0$ and $1$.
While the radial parts of the wave functions for different $M_J$ 
are identical, the angular parts differ.
Specifically, given the scattering wave function in 
the $M_J$ channel,
that in the $M_J^{\prime}$ channel
is obtained by replacing 
$|J,M_J;L,S\rangle$ by $|J,M_J^{\prime};L,S\rangle$.
Angular momentum algebra yields that the components
$(J,M_J;L,S)=(1,M_J;1,0)$, $(1,M_J;1,1)$, $(1,M_J;2,1)$ and $(1,M_J;0,1)$ contribute.
For the $(J,M_J)=(1,M_J)$ channels,
there exist two scattering thresholds, namely, $E=-\eta^2 E_r$ and
$E=-E_r$.
For $E> -\eta^2E_r$, branch A and branch B are open.
At $E=-\eta^2 E_r$, branch B becomes closed,
implying that branch A is the only open branch in
the energy regime $-E_r<E< -\eta^2 E_r$ (see the dispersion curves in Fig.~\ref{fig1}).
Using the projection procedure outlined in Sec.~\ref{sec2}, 
we have
\begin{widetext}
\begin{eqnarray}
\label{mathcal_J_1M_J}
\mathcal{J}^{(1,M_J)}(r)=
\begin{bmatrix}
&\frac{\imath}{\sqrt{2}}N_{A_+}\bar{k}_{A_+}j_1(\bar{k}_{A_+}r)
&\frac{\imath}{\sqrt{2}}N_{A_-}\bar{k}_{A_-}j_1(\bar{k}_{A_-}r)
&0
&0\\
&\frac{1}{\sqrt{6}}N_{A_+}\bar{k}_{A_+}j_0(\bar{k}_{A_+}r)
&\frac{1}{\sqrt{6}}N_{A_-}\bar{k}_{A_-}j_0(\bar{k}_{A_-}r) 
&\frac{1}{\sqrt{3}}N_{B_+}\bar{k}_{B_+}j_0(\bar{k}_{B_+}r)
&\frac{1}{\sqrt{3}}N_{B_-}\bar{k}_{B_-}j_0(\bar{k}_{B_-}r)\\
&\frac{1}{\sqrt{3}}N_{A_+}\bar{k}_{A_+}j_{2}(\bar{k}_{A_+}r)
&\frac{1}{\sqrt{3}}N_{A_-}\bar{k}_{A_-}j_{2}(\bar{k}_{A_-}r)
&-\frac{1}{\sqrt{6}}N_{B_+}\bar{k}_{B_+}j_{2}(\bar{k}_{B_+}r)
&-\frac{1}{\sqrt{6}}N_{B_-}\bar{k}_{B_-}j_{2}(\bar{k}_{B_-}r)\\
&0
&0
&\frac{-\imath}{\sqrt{2}}N_{B_+}\bar{k}_{B_+}j_{1}(\bar{k}_{B_+}r)
&\frac{-\imath}{\sqrt{2}}N_{B_-}\bar{k}_{B_-}j_{1}(\bar{k}_{B_-}r)
\end{bmatrix},
\end{eqnarray}
\begin{eqnarray}
\label{mathcal_N_1M_J}
\mathcal{N}^{(1,M_J)}(r)=
\begin{bmatrix}
&\frac{\imath}{\sqrt{2}}N_{A_+}\bar{k}_{A_+}n_1(\bar{k}_{A_+}r)
&-\frac{\imath}{\sqrt{2}}N_{A_-}\bar{k}_{A_-}n_1(\bar{k}_{A_-}r)
&0
&0\\
&\frac{1}{\sqrt{6}}N_{A_+}\bar{k}_{A_+}n_0(\bar{k}_{A_+}r)
&-\frac{1}{\sqrt{6}}N_{A_-}\bar{k}_{A_-}n_0(\bar{k}_{A_-}r) 
&\frac{1}{\sqrt{3}}N_{B_+}\bar{k}_{B_+}n_0(\bar{k}_{B_+}r)
&-\frac{1}{\sqrt{3}}N_{B_-}\bar{k}_{B_-}n_0(\bar{k}_{B_-}r)\\
&\frac{1}{\sqrt{3}}N_{A_+}\bar{k}_{A_+}n_{2}(\bar{k}_{A_+}r)
&-\frac{1}{\sqrt{3}}N_{A_-}\bar{k}_{A_-}n_{2}(\bar{k}_{A_-}r)
&-\frac{1}{\sqrt{6}}N_{B_+}\bar{k}_{B_+}n_{2}(\bar{k}_{B_+}r)
&\frac{1}{\sqrt{6}}N_{B_-}\bar{k}_{B_-}n_{2}(\bar{k}_{B_-}r)\\
&0
&0
&\frac{-\imath}{\sqrt{2}}N_{B_+}\bar{k}_{B_+}n_{1}(\bar{k}_{B_+}r)
&\frac{\imath}{\sqrt{2}}N_{B_-}\bar{k}_{B_-}n_{1}(\bar{k}_{B_-}r)
\end{bmatrix},
\end{eqnarray}
and
\begin{eqnarray}
\label{wave_function_1M_j}
\Psi^{(1,M_J)}(r)=\mathcal{J}^{(1,M_J)}(r)-\mathcal{N}^{(1,M_J)}(r)\mathcal{K}^{(1,M_J)}.
\end{eqnarray}

Assuming $E > -\eta^2 E_r$
and following the same steps as in Sec.~\ref{sec3},
the $4$ by $4$ $\mathcal{K}^{(1,M_J)}$ matrix is determined by matching the zero-range 
boundary conditions.
We find 
\begin{eqnarray}
\label{Kmatrix_J=1}
\mathcal{K}^{(1,M_J)}=-\frac{a_s}{3(\bar{k}_{A_+}-\bar{k}_{A_-})}
\begin{bmatrix}
\bar{k}_{A_+}^2 & \bar{k}_{A_+}\bar{k}_{A_-} &\beta \bar{k}_{A_+}\bar{k}_{B_+} &\beta \bar{k}_{A_+}\bar{k}_{B_-}\\
\bar{k}_{A_+}\bar{k}_{A_-} & \bar{k}_{A_-}^2 &\beta \bar{k}_{A_-}\bar{k}_{B_+} &\beta \bar{k}_{A_-}\bar{k}_{B_-}\\
\beta \bar{k}_{A_+}\bar{k}_{B_+} &\beta \bar{k}_{A_-}\bar{k}_{B_+}  &\beta^2\bar{k}_{B_+}^2 &\beta^2\bar{k}_{B_+}\bar{k}_{B_-}\\
\beta \bar{k}_{A_+}\bar{k}_{B_-} &\beta \bar{k}_{A_-}\bar{k}_{B_-} &\beta^2\bar{k}_{B_+}\bar{k}_{B_-} &\beta^2\bar{k}_{B_-}^2 
\end{bmatrix}
,
\end{eqnarray}
\end{widetext}
where
$\beta$ is equal to $\sqrt{2}N_{B_+}/N_{A_+}$. 
The energy-dependent $\mathcal{S}^{(1,M_J)}$ matrix
and partial cross sections 
$\sigma_{jl}^{(1,M_J)}$, 
in turn, 
are obtained using Eqs.~\eqref{mathcal_S} and \eqref{partial_cross_section}.
Note, throughout the remainder of this section
we drop the superscript ``$(1,M_J)$'' from the partial cross sections
$\sigma_{jl}^{(1,M_J)}$ for notational convenience.

To obtain the partial cross sections for the
energy region $-E_r<E<-\eta^2E_r$,
we pursue two different but equivalent approaches. 
Approach 1 constructs a $4$ by $4$ $\mathcal{K}^{(1,M_J)}$ matrix,
partitions off the $2$ by $2$ $\mathcal{K}^{(1,M_J)}_{oo}$
matrix, 
and then constructs
the $2$ by $2$ $\mathcal{S}^{(1,M_J)}_{oo}$ matrix.
Approach 2 analytically continues
the $4$ by $4$ $\mathcal{S}^{(1,M_J)}$ matrix for $E> -\eta^2 E_r$
to the $E<-\eta^2 E_r$ energy region and 
then partitions off the $2$ by $2$ $\mathcal{S}^{(1,M_J)}_{oo}$
matrix. In both approaches, the partial cross sections are obtained
from
the $2$ by $2$ $\mathcal{S}^{(1,M_J)}_{oo}$ matrix.

Approach 1 follows the logic of Ref.~\cite{johnson}, which discusses
multi-channel scattering in the absence of spin-orbit coupling. 
Since branch $B$ is closed in the energy region $-E_r<E<-\eta^2E_r$,
$\bar{k}_{B_+}$ and $\bar{k}_{B_-}$ are imaginary.
This motivates us to
write 
$\bar{k}_{B_+}=\imath k^{*}-\eta k_{\text{so}}$ 
and $\bar{k}_{B_-}=-\imath k^{*}-\eta k_{\text{so}}$ with $k^*$
real and 
$k^*$ greater than zero.
To ensure that  
the last two columns of $\mathcal{N}^{(1,M_J)}(r)$ decay exponentially,
we replace the $n_{j}(\bar{k}_{B_+}r)$ ($j=0,1$, and $2$) 
in the third column of $\mathcal{N}^{(1,M_J)}(r)$ by 
$-n_{j}(\bar{k}_{B_+}r)+\imath j_{j}(\bar{k}_{B_+}r)$ 
and
the $n_{j}(\bar{k}_{B_-}r)$  ($j=0,1$, and $2$) in the fourth
column of $\mathcal{N}^{(1,M_J)}(r)$ by
 $-n_{j}(\bar{k}_{B-}r)-\imath j_{j}(\bar{k}_{B_-}r)$.
To determine the corresponding $\mathcal{K}^{(1,M_J)}$ matrix,
we plug $\mathcal{J}^{(1,M_J)}(r)$,
Eq.~(\ref{mathcal_J_1M_J}), and the modified $\mathcal{N}^{(1,M_J)}(r)$
into Eq.~\eqref{wave_function_1M_j}.
Matching the resulting 
$\Psi^{(1,M_J)}(r)$ to the $s$-, $p$- and $d$-wave boundary conditions
at 
$r=0$,
we obtain a modified $4$ by $4$ $\mathcal{K}^{(1,M_J)}$ matrix 
applicable to the energy regime $-E_r<E<-\eta^2E_r$.
Dividing this matrix into
the open-open sub-block $\mathcal{K}^{(1,M_J)}_{oo}$,
the closed-closed sub-block $\mathcal{K}^{(1,M_J)}_{cc}$,
the open-closed sub-block $\mathcal{K}^{(1,M_J)}_{oc}$,
and the closed-open sub-block $\mathcal{K}^{(1,M_J)}_{co}$,
we have
\begin{eqnarray}
\label{K_matrix_openclosed}
\mathcal{K}^{(1,M_J)}=
\begin{bmatrix}
&\mathcal{K}^{(1,M_J)}_{oo} &\mathcal{K}^{(1,M_J)}_{oc}\\
&\mathcal{K}^{(1,M_J)}_{co} &\mathcal{K}^{(1,M_J)}_{cc}  
\end{bmatrix}
,
\end{eqnarray}
where the open-open block
$\mathcal{K}^{(1,M_J)}_{oo}$ reads
\begin{eqnarray}
  \label{K_matrix_oo}
  \mathcal{K}^{(1,M_J)}_{oo}=\frac{-a_{\text{eff}}}{\bar{k}_{A_+}-\bar{k}_{A_-}}
  \begin{bmatrix}
    \bar{k}_{A_+}^2 & \bar{k}_{A_+}\bar{k}_{A_-}\\
    \bar{k}_{A_+}\bar{k}_{A_-} & \bar{k}_{A_-}^2
    \end{bmatrix}
\end{eqnarray}
with
\begin{eqnarray}
  \label{a_eff}
  a_{\text{eff}}=\frac{a_sk^*}{3k^*+2a_s\left[(\eta k_{\text{so}})^2-(k^*)^2\right]}
.
  \end{eqnarray}
Interestingly,
Eq.~\eqref{K_matrix_oo} is identical
to Eq.~\eqref{mathcal_K_00}
with $a_s$ replaced by $a_{\text{eff}}$.
The quantity $a_{\text{eff}}$ can thus be interpreted as an 
effective scattering length, which is modified by the states in the 
closed branch B, that describes the scattering between the
states in branch A. 
The $\mathcal{S}^{(1,M_J)}_{oo}$ matrix 
is obtained using Eq.~\eqref{mathcal_S}
with $\mathcal{K}^{(J,M_J)}$ replaced by
$\mathcal{K}^{(J,M_J)}_{oo}$.
Last,
the partial cross sections 
$\sigma_{11}$, $\sigma_{12}$, $\sigma_{21}$, 
and $\sigma_{22}$ for $-E_r\le E<-\eta^2E_r$ 
are obtained using Eq.~\eqref{partial_cross_section} 
with $\mathcal{S}^{(J,M_J)}$ replaced by $\mathcal{S}^{(1,M_J)}_{oo}$.

Approach 2 is based on analytic continuation. While this
approach may be more intuitive to some readers, we note that the
approach may be impractical for numerical 
calculations
that propagate
the logarithmic derivative matrix.
Approach 2 takes the 
$4$ by $4$ $\mathcal{S}^{(1,M_J)}$
matrix for $E> -\eta^2 E_r$ and 
replaces
$\bar{k}_{B_+}$
and
$\bar{k}_{B_-}$
by (this is the same as in approach 1), respectively, 
$\imath k^{*}-\eta k_{\text{so}}$ 
and 
$-\imath k^{*}-\eta k_{\text{so}}$, where $k^*$ is
real and greater than zero.
These replacements guarantee that 
the outgoing solutions $\psi_B^{(1,M_J)+}(\bar{k}_{B_+}r)$ and $\psi_B^{(1,M_J)-}(\bar{k}_{B_-}r)$
decay exponentially for $E < -\eta^2E_r$.
We then divide 
$\mathcal{S}^{(1,M_J)}$ into the 
open-open sub-block $\mathcal{S}^{(1,M_J)}_{oo}$,
the closed-closed sub-block $\mathcal{S}^{(1,M_J)}_{cc}$,
the open-closed sub-block $\mathcal{S}^{(1,M_J)}_{oc}$,
and the closed-open sub-block $\mathcal{S}^{(1,M_J)}_{co}$,
\begin{eqnarray}
\label{S_matrix_openclosed}
\mathcal{S}^{(1,M_J)}=
\begin{bmatrix}
&\mathcal{S}^{(1,M_J)}_{oo} &\mathcal{S}^{(1,M_J)}_{oc}\\
&\mathcal{S}^{(1,M_J)}_{co} &\mathcal{S}^{(1,M_J)}_{cc}  
\end{bmatrix}
.
\end{eqnarray}
The $\mathcal{S}^{(1,M_J)}_{oo}$ matrix obtained from approach 2
agrees, as it should, with that obtained from approach 1.

\begin{figure}
\vspace*{0.8cm}
\hspace*{0cm}
\includegraphics[width=0.4\textwidth]{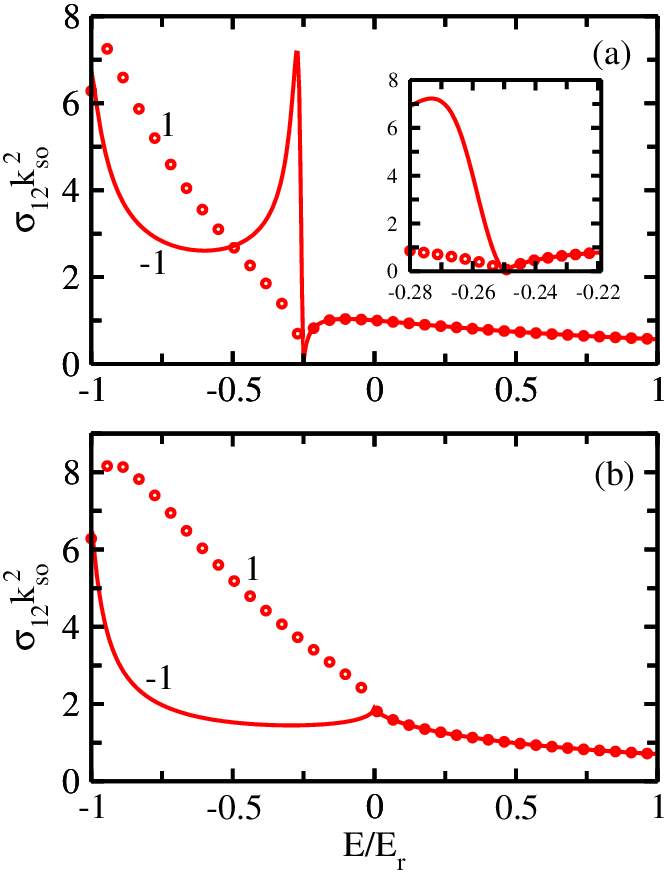}
\caption{(Color online) 
Partial cross section 
$\sigma_{12}=\sigma_{22}$,
in units of $1/k_{\text{so}}^2$,  
for the $(J,M_J)=(1,M_J)$ channels as a function of the energy $E$
for (a) unequal masses 
($\eta=1/2$) 
and (b) equal masses ($\eta=0$);
this figure considers scattering processes within branch A.
The red solid curves correspond to $a_s k_{\text{so}}=-1$
and the 
circles
to $a_s k_{\text{so}}=1$.
For $E\ge -\eta^2E_r$,
the 
partial
cross section
is
independent of the sign of
the $s$-wave scattering length,
i.e., the solid 
curves and the circles
coincide.
For $-E_r\le E<-\eta^2E_r$, in contrast, the 
partial 
cross 
section depends
on the sign of $a_s$.
The inset in panel (a) shows
an enlargement of the $E\approx -\eta^2E_r$ region,
demonstrating that the partial cross section changes 
continuously with energy.
}
\label{fig3a}
\end{figure}

We first analyze the scattering between branch A states,
i.e., we analyze 
the partial cross sections $\sigma_{jl}$ with
$j,l$ equal to $1$ and $2$.
The threshold behavior is obtained
by Taylor expanding $\sigma_{11}=\sigma_{21}$ and $\sigma_{12}=\sigma_{22}$ 
around 
$E=-\eta^2E_r$ 
(this is the energy where branch B becomes
closed) 
and $E=-E_r$ (this is the lowest scattering threshold).
Table~\ref{threshold_11_12} shows the leading and sub-leading terms
for $\eta\ne 0$ (columns 2-4) and $\eta=0$ (columns 5-7).
\begin{table*}
\begin{center}
\begin{tabular}{c|ccc|ccc}
 & $E \to (-\eta^2 E_r)^+$ & $E \to (-\eta^2 E_r)^-$ &  $E \to - E_r$ & $E \to 0^+$ & $E \to 0^-$ & $E \to -E_r$ \\ 
 & $(\eta, a_s \ne 0)$ & $(\eta, a_s \ne 0)$ &  $(\eta, a_s \ne 0)$ & $(\eta=0, a_s \ne 0)$ & $(\eta=0, a_s \ne 0)$ & ($\eta=0$, $a_s \ne 0$) \\  \hline
$ \sigma_{11} k_{\text{so}}^2 = \sigma_{21} k_{\text{so}}^2$ & 
$c_{11}^{(0)} y^2+c_{11}^{(1+)}y^3$ & 
$c_{11}^{(0)} y^2+{c_{11}^{(1-)}y^3}/{(a_s k_{\text{so}})}$ & 
$2 \pi - 4 \pi x$ & $d_{11}^{(0)} y^4+d_{11}^{(1)}y^5$ & 
$d_{11}^{(0)} y^4+{3d_{11}^{(1)} y^5}/{(2 a_s k_{\text{so}})}$ &
$2 \pi - 4 \pi x$ \\
$ \sigma_{12} k_{\text{so}}^2 = \sigma_{22} k_{\text{so}}^2$ & 
$c_{12}^{(0)} y^2+c_{12}^{(1+)}y^3$ & 
$c_{12}^{(0)} y^2+{c_{12}^{(1-)}y^3}/{(a_s k_{\text{so}})}$ & 
$2 \pi + 4 \pi x$ & $d_{12}^{(0)} +d_{12}^{(1)}y$ & 
$d_{12}^{(0)} +{3d_{12}^{(1)} y}/{(2 a_s k_{\text{so}})}$ &
$2 \pi + 4 \pi x$ \\
\end{tabular}
\caption{Behavior of the partial cross sections 
for two particles
with unequal 
masses ($\eta\ne 0$) and 
equal masses ($\eta=0$) in the $(J,M_J)=(1,M_J)$ channels;
this table considers scattering processes within branch A.
The superscript ``$+$'' indicates that the limit is taken from above
(positive side) 
while the superscript ``$-$'' indicates that it is 
taken from below
(negative side).
The coefficients $c_{11}^{(0)}$, $c_{11}^{(1+)}$, $c_{11}^{(1-)}$, $c_{12}^{(0)}$, 
$c_{12}^{(1+)}$, and $c_{12}^{(1-)}$
for $\eta\ne 0$,
which are reported in Appendix~\ref{appendixB},
are independent of $a_s$.
In contrast,
the coefficients $d_{11}^{(0)}$, $d_{11}^{(1)}$, 
$d_{12}^{(0)}$, and $d_{12}^{(1)}$
for $\eta=0$, which are also reported in Appendix~\ref{appendixB},
depend on $a_s^2$.
The quantity $x$ is defined in the caption of 
Table~\ref{table:threshold_00} and $y$ is equal to
$\sqrt{|E/E_r+\eta^2|}$.
}
\label{threshold_11_12}
\end{center}
\end{table*}
It can be seen that the sub-leading term of $\sigma_{jl}$
behaves differently
for $E \to (-\eta^2E_r)^+$ and $E \to (-\eta^2E_r)^-$.
This ``asymmetric'' behavior is exemplarily shown in 
Fig.~\ref{fig3a} for $\sigma_{12}k_{\text{so}}^2$ for
$a_s k_{\text{so}} = 1$ 
(circles)
and
$a_s k_{\text{so}} = -1$ 
(solid lines).
The scattering threshold is equal to $-E_r/4$
for $\eta=1/2$ [see Fig.~\ref{fig3a}(a)] and
equal to $0$
for $\eta=0$ [see Fig.~\ref{fig3a}(b)].
Figure~\ref{fig3a} also illustrates another important
aspect.
The partial wave cross section $\sigma_{12}$ is independent of 
the sign of the scattering length $a_s$ 
for $E>-\eta^2 E$ if $\eta$ is finite
and
for $E>0$ if $\eta$ is zero.
The sign of the scattering length does, however, enter into the
partial cross section expressions below these energies.
The same holds true for $\sigma_{11}=\sigma_{21}$.
This results from the fact that the partial cross sections
$\sigma_{jl}$ ($j,l=1$ and 2)
depend on $a_s^2$ for $E > -\eta^2 E_r$ and on 
$a_{\text{eff}}^2$ for $E< -\eta^2 E_r$.
While
$a_s^2$ is independent of the sign of
$a_s$,
$a_{\text{eff}}^2$ depends on the sign of $a_s$ [see Eq.~(\ref{a_eff})].
This behavior should be contrasted with the usual $s$-wave case, where the 
cross section is independent of the sign of the $s$-wave scattering length
for all positive energies,
implying that cross section measurements can only be used to deduce the
magnitude of the scattering length but not the sign.
Another aspect 
illustrated by Fig.~\ref{fig3a} 
is that 
$\sigma_{12}$ vanishes at $E=-\eta^2E_r$ for $\eta=1/2$ and is finite for $\eta=0$.
This 
can be
understood by evaluating  
$a_{\text{eff}}$,
Eq.~(\ref{a_eff}),  
for
$E=-\eta^2E_r$ ($k^*=0$).
For $\eta \ne 0$ and $E=-\eta^2E_r$,
$a_{\text{eff}}$ is equal
to zero, implying that the partial cross sections
for branch A vanish at this energy.
For $\eta=0$ and $E=0$, $a_{\text{eff}}$ is equal to 
$a_s/3$, 
implying that the partial cross sections for
branch A are finite.
The partial wave cross sections $\sigma_{jl}$ with $j,l$ equal to
1 and 2 approach 
$2\pi/k_{\text{so}}^2$ at the lowest scattering threshold, i.e., at 
$E=-E_r$
(see Fig.~\ref{fig3a} and columns 4 and 7 of Table~\ref{threshold_11_12}).
Importantly, 
the sub-leading term is---as in the $(J,M_J)=(0,0)$ channel---independent of 
$a_s$
($a_s \ne 0$) for vanishing and finite
$\eta$. This can be interpreted, as in Sec.~\ref{sec3},
as a consequence of an effective dimensionality reduction.

\begin{figure}
\vspace*{0.5cm}
\hspace*{0cm}
\includegraphics[width=0.45\textwidth]{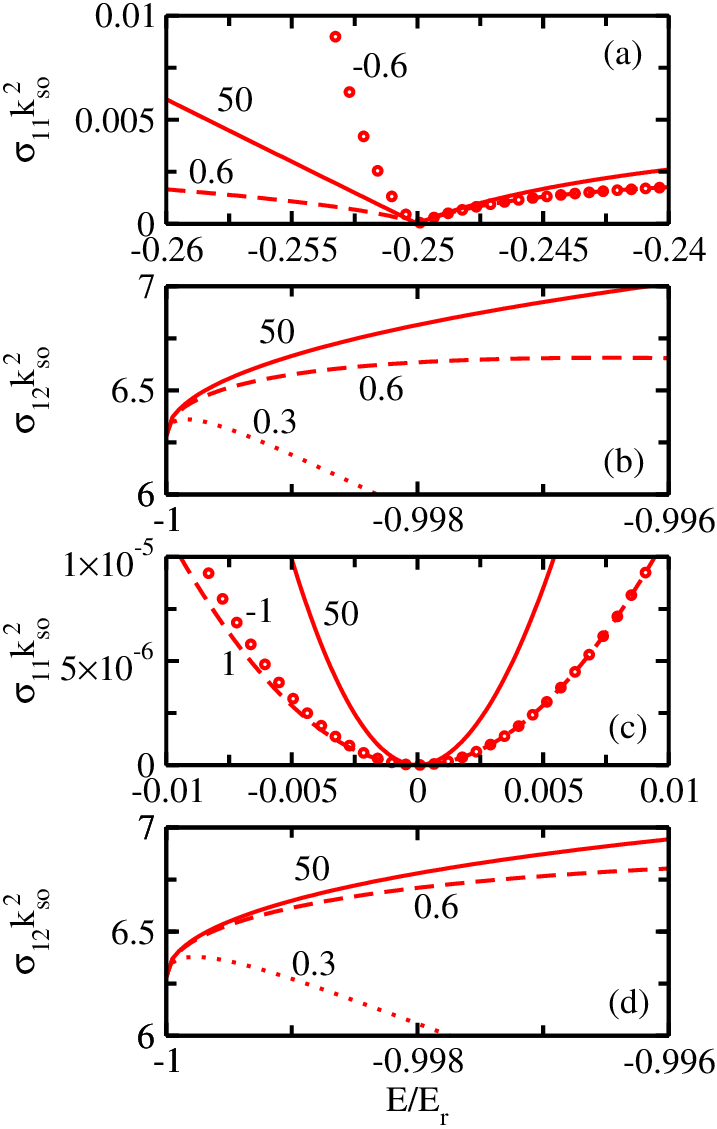}
\caption{(Color online)
Partial cross sections,
in units of $1/k_{\text{so}}^2$,  
for the $(J,M_J)=(1,M_J)$ channels as a function of the energy $E$
for (a) and (b) unequal masses 
$(\eta=1/2)$, 
and 
(c) and (d) equal masses ($\eta=0$);
this figure considers scattering processes within branch A.
Panel~(a) shows 
$\sigma_{11}k_{\text{so}}^2=\sigma_{21}k_{\text{so}}^2$ 
in the vicinity of $E=-\eta^2E_r$
for $a_sk_{\text{so}}=50$
(solid line), 
$a_sk_{\text{so}}=0.6$ (dashed line), and $a_sk_{\text{so}}=-0.6$ 
(circles).
Panel~(b) shows 
$\sigma_{12}k_{\text{so}}^2=\sigma_{22}k_{\text{so}}^2$ 
in the vicinity of $E=-E_r$
for $a_sk_{\text{so}}=50$
(solid line), 
$a_sk_{\text{so}}=0.6$ (dashed line), and $a_sk_{\text{so}}=0.3$ (dotted line).
Panel~(c) shows 
$\sigma_{11}k_{\text{so}}^2=\sigma_{21}k_{\text{so}}^2$ 
in the vicinity of $E=0$
for $a_sk_{\text{so}}=50$
(solid line), 
$a_sk_{\text{so}}=1$ (dashed line), and $a_sk_{\text{so}}=-1$ 
(circles).
Panel~(d) shows 
$\sigma_{12}k_{\text{so}}^2=\sigma_{22}k_{\text{so}}^2$ 
in the vicinity of $E=-E_r$
for $a_sk_{\text{so}}=50$
(solid line), 
$a_sk_{\text{so}}=0.6$ (dashed line), and $a_sk_{\text{so}}=0.3$ (dotted line).
}
\label{fig3b}
\end{figure}

Figure~\ref{fig3b} highlights a number of other characteristics
of the partial cross sections $\sigma_{12}$ and $\sigma_{11}$ in the vicinity of the threshold
energies.
For $\eta \ne 0$, 
the leading order term of $\sigma_{11}$ in the vicinity of $E=-\eta^2E_r$
goes to zero linearly with $|E+\eta^2E_r|$, 
with a coefficient
that is independent of $a_s$.
The sub-leading term of $\sigma_{11}$ scales as
$|E+\eta^2 E_r|^{3/2}$ and is independent of $a_s$
for $E\to(-\eta^2E_r)^{+}$ but
dependent on $a_s$ for $E\to(-\eta^2E_r)^{-}$.
This implies that 
$\sigma_{11}$ depends very weakly on $a_s$
for $E\to(-\eta^2E_r)^{+}$ and comparatively strongly on
$a_s$ 
for $E\to(-\eta^2E_r)^-$.
This discussion explains the asymmetry of the partial cross section $\sigma_{11}$ 
in the vicinity of
$E=-\eta^2 E_r$
in Fig.~\ref{fig3b}(a).

For $\eta = 0$, in contrast, 
the leading order term of $\sigma_{11}$ in the vicinity of $E=0$
goes to zero quadratically with $E$, with a coefficient
that depends on $a_s$ for $E<0$ and on $|a_s|$ for $E>0$
[see Fig.~\ref{fig3b}(c)].
Figures~\ref{fig3b}(b) and \ref{fig3b}(d) show the threshold behavior 
of
$\sigma_{12}$
for $\eta=1/2$ and $\eta=0$, respectively,
in the vicinity of $E=-E_r$
for
$a_sk_{\text{so}}=50$, $0.6$ and $0.3$
(see the labels in the figure).
These figures illustrate the dependence of the partial cross section 
on $a_s k_{\text{so}}$.

\begin{figure}
\vspace*{0.5cm}
\hspace*{0cm}
\includegraphics[width=0.4\textwidth]{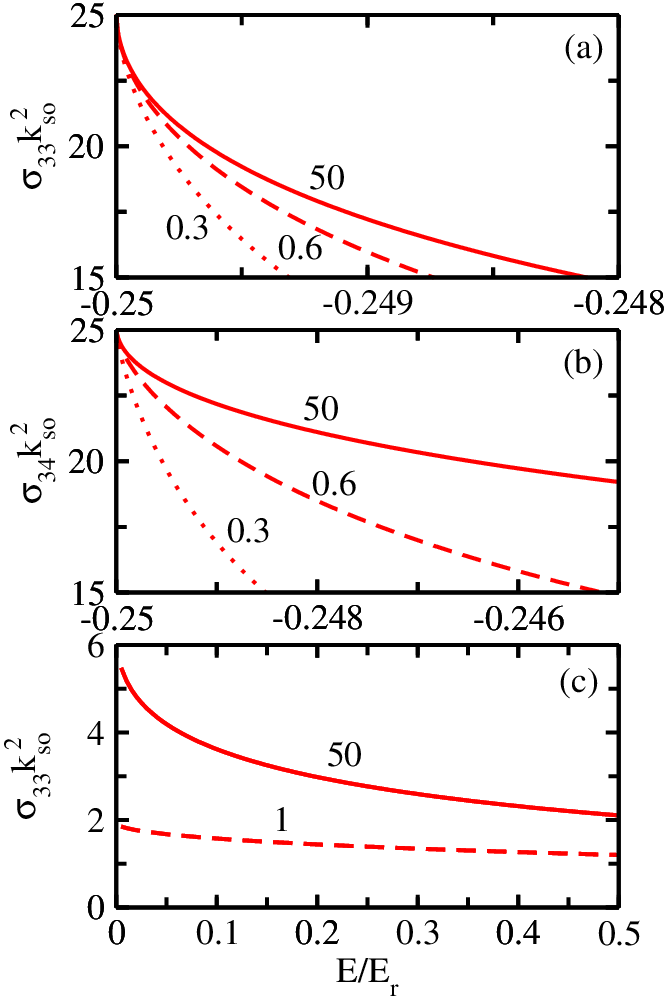}
\caption{(Color online)
Partial cross sections,
in units of $1/k_{\text{so}}^2$,  
for the $(J,M_J)=(1,M_J)$ channels as a function of the energy $E$
for (a) and (b) unequal masses 
$(\eta=1/2)$, 
and 
(c) equal masses ($\eta=0$);
this figure considers scattering processes within branch B.
Panels~(a) and (b) show
$\sigma_{33}k_{\text{so}}^2=\sigma_{43}k_{\text{so}}^2$ 
and
$\sigma_{34}k_{\text{so}}^2=\sigma_{44}k_{\text{so}}^2$, 
respectively,
in the vicinity of $E=-\eta^2E_r$
for $|a_sk_{\text{so}}|=50$
(solid line), 
$|a_sk_{\text{so}}|=0.6$ (dashed line), and $|a_sk_{\text{so}}|=0.3$ (dotted line).
Panel~(c) shows 
$\sigma_{33}k_{\text{so}}^2=\sigma_{34}k_{\text{so}}^2=\sigma_{43}k_{\text{so}}^2=\sigma_{44}k_{\text{so}}^2$ 
in the vicinity of $E=0$
for $|a_sk_{\text{so}}|=50$
(solid line) and 
$|a_sk_{\text{so}}|=1$ (dashed line).
}
\label{fig4}
\end{figure}

Next,
we 
analyze
the partial cross sections $\sigma_{33}$, $\sigma_{34}$, $\sigma_{43}$ and $\sigma_{44}$,
which
describe
the scattering between states in branch B.
We Taylor expand 
$\sigma_{33}=\sigma_{34}$ and $\sigma_{43}=\sigma_{44}$
for $\eta>0$
around $E=-\eta^2E_r$
(see column 2 of Table~\ref{table:threshold_1M_J_33_34}).
Since 
branch B is closed
below $E=-\eta^2E_r$,
the Taylor series 
is
only 
applicable to the energy regime $E \ge -\eta^2E_r$.
For $\eta\ne 0$, 
$\sigma_{33}=\sigma_{34}$ and $\sigma_{43}=\sigma_{44}$
approach 
the
constant $2\pi/(\eta^2k_{\text{so}}^2)$
(at this threshold all other partial cross sections go to
zero, provided $\eta$ is greater than $0$;
see Tables~\ref{threshold_11_12}-\ref{table:threshold_1M_J_31_41}).
Figures~\ref{fig4}(a) and \ref{fig4}(b)
show 
$\sigma_{33}$ and $\sigma_{34}$,
respectively, 
in the vicinity of $E = -\eta^2E_r$
for $\eta=1/2$ and three
different $|a_sk_{\text{so}}|$ combinations, i.e., for
 $|a_sk_{\text{so}}|=50$, $0.6$, and $0.3$.
To obtain the threshold behavior for $\eta=0$,
we set $\eta$ to $0$ and Taylor expand 
$\sigma_{33}=\sigma_{34}$ and $\sigma_{43}=\sigma_{44}$
around $E=0$.
The resulting leading-order  
terms
(see column 3 of Table~\ref{table:threshold_1M_J_33_34}) 
are equal to each other and
depend on
$a_s^2$.
\begin{table}
\begin{center}
\begin{tabular}{c|c|c}
  &$E\to(-\eta^2E_r)^{+}$ &$E\to 0^{+}$\\
  &$(\eta, a_s\ne 0)$ &
$(\eta=0,a_s\ne 0)$\\
\hline
$\sigma_{33}k_{\text{so}}^2=\sigma_{43}k_{\text{so}}^2$
&${2\pi}/{\eta^2}$ 
&${8\pi a_s^2k_{\text{so}}^2}/{(9+4a_s^2k_{\text{so}}^2)}$\\
$\sigma_{34}k_{\text{so}}^2=\sigma_{44}k_{\text{so}}^2$
&${2\pi}/{\eta^2}$ 
&${8\pi a_s^2k_{\text{so}}^2}/{(9+4a_s^2k_{\text{so}}^2)}$\\
\end{tabular}
\caption{
Behavior of the partial cross sections 
for two particles
with unequal masses 
($\eta\ne 0$) and equal masses ($\eta=0$) in the $(J,M_J)=(1,M_J)$ channels
in the vicinity of the energy where branch B becomes closed;
this table considers scattering processes within branch B.
The leading term
for $\eta \ne 0$ 
is constant
(independent of $a_s$).
In contrast,
the leading term
for $\eta=0$
depends 
on $a_s^2$.
}
\label{table:threshold_1M_J_33_34}
\end{center}
\end{table}
In fact, all four partial cross sections $\sigma_{33}$, $\sigma_{43}$,
$\sigma_{34}$, and $\sigma_{44}$ coincide for
all positive energies. 
Figure~\ref{fig4}(c) shows $\sigma_{33}=\sigma_{34}$
for $\eta=0$ with $|a_sk_{\text{so}}|=50$ and $1$.

\begin{figure}
\vspace*{1cm}
\hspace*{0cm}
\includegraphics[width=0.45\textwidth]{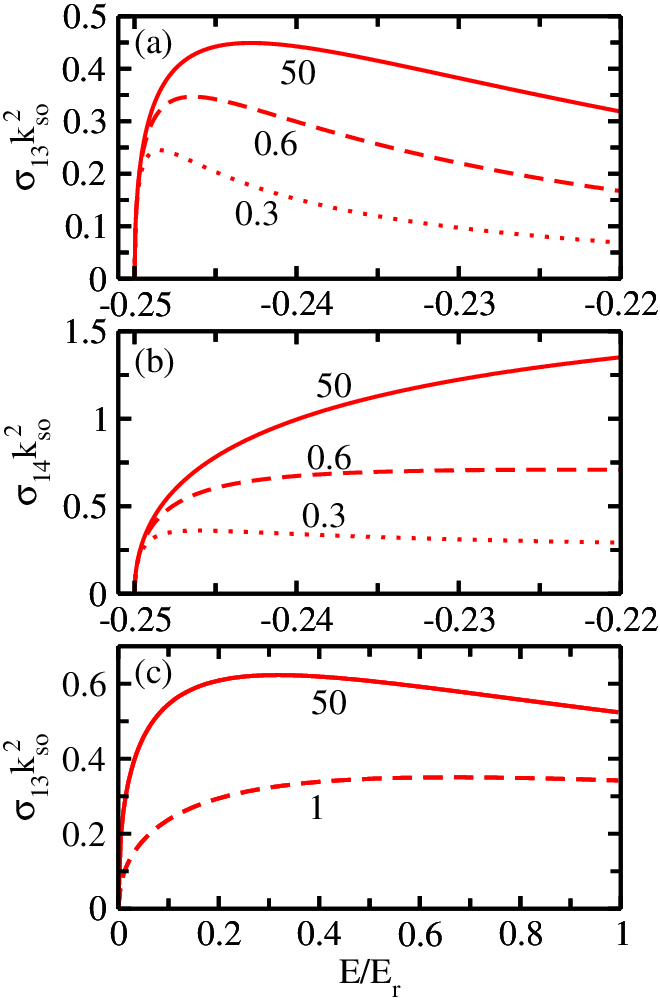}
\caption{(Color online)
Partial cross sections,
in units of $1/k_{\text{so}}^2$,  
for the $(J,M_J)=(1,M_J)$ channels as a function of the energy $E$
for (a) and (b) unequal masses 
$(\eta=1/2)$, 
and 
(c) equal masses ($\eta=0$);
this figure considers scattering processes from branch A to branch B.
Panel~(a) shows 
$\sigma_{13}k_{\text{so}}^2 = \sigma_{23}k_{\text{so}}^2$ 
in the vicinity of $E=-\eta^2E_r$
for $a_sk_{\text{so}}=50$
(solid line), 
$a_sk_{\text{so}}=0.6$ (dashed line), and $a_sk_{\text{so}}=0.3$ (dotted line).
Panel~(b) shows 
$\sigma_{14}k_{\text{so}}^2 = \sigma_{24}k_{\text{so}}^2$ 
in the vicinity of $E=-\eta^2E_r$
for $a_sk_{\text{so}}=50$
(solid line), 
$a_sk_{\text{so}}=0.6$ (dashed line), and $a_sk_{\text{so}}=0.3$ (dotted line).
Panel~(c) shows 
$\sigma_{13}k_{\text{so}}^2=\sigma_{23}k_{\text{so}}^2=\sigma_{14}k_{\text{so}}^2=\sigma_{24}k_{\text{so}}^2$ 
in the vicinity of $E=0$
for $|a_sk_{\text{so}}|=50$
(solid line) and 
$|a_sk_{\text{so}}|=1$ (dashed line).
}
\label{fig5}
\end{figure}

Next, we 
analyze
the partial cross sections 
$\sigma_{13}$, $\sigma_{14}$, $\sigma_{23}$ and $\sigma_{24}$,
which describe
scattering 
processes from states in branch A to states in branch B.
Taylor expanding $\sigma_{13}=\sigma_{23}$ and $\sigma_{14}=\sigma_{24}$
around $E=-\eta^2E_r$,
we find that the leading order term---assuming $E \ge -\eta^2 E_r$---is
proportional to
$\sqrt{|E+\eta^2E_r|}$
for $\eta \ne 0$ and $\eta=0$
(see Table~\ref{table:threshold_1M_J_13_14}).
\begin{table}
\begin{center}
\begin{tabular}{c|c|c}
  &$E\to(-\eta^2E_r)^{+}$ &$E\to 0^{+}$\\
  &$(\eta, a_s \ne 0)$ &$(\eta=0,a_s \ne 0)$\\
  \hline
  $\sigma_{13}k_{\text{so}}^2=\sigma_{23}k_{\text{so}}^2$
  &$c_{13}^{(0)}y$ &$d_{13}^{(0)}y$\\
  $\sigma_{14}k_{\text{so}}^2=\sigma_{24}k_{\text{so}}^2$
  &$c_{14}^{(0)}y$ &$d_{14}^{(0)}y$
\end{tabular}
\caption{Behavior of the partial cross sections 
for two particles
with unequal masses ($\eta\ne 0$) and equal masses ($\eta=0$) in the $(J,M_J)=(1,M_J)$ channels
in the vicinity of the energy where branch B becomes closed;
this table considers scattering processes from branch A to branch B.
The coefficients $c_{13}^{(0)}$ and $c_{14}^{(0)}$
for $\eta\ne 0$,
which are reported in Appendix~\ref{appendixB},
are independent of $a_s$.
In contrast,
the coefficients $d_{13}^{(0)}$ and $d_{14}^{(0)}$
for $\eta=0$, which are also reported in Appendix~\ref{appendixB},
depend on $a_s^2$.
}
\label{table:threshold_1M_J_13_14}
\end{center}
\end{table}
The prefactor is independent of $a_s$ for $\eta\ne 0$ 
and depends quadratically on $a_s$ for $\eta=0$.
Figures~\ref{fig5}(a) and \ref{fig5}(b)
illustrate the threshold behavior of the partial cross sections
$\sigma_{13}$ and $\sigma_{14}$, respectively, for $\eta=1/2$ 
and three different $|a_sk_{\text{so}}|$
combinations, i.e., for $|a_sk_{\text{so}}|=50$, $0.6$, and $0.3$.
Close to $-\eta^2E_r$, the leading term dominates
and all three curves coincide approximately.
Figure~\ref{fig5}(c) 
shows $\sigma_{13}=\sigma_{14}$
for $\eta=0$ and $|a_sk_{\text{so}}|=50$ (solid line)
and $|a_sk_{\text{so}}|=1$ (dashed line).
The partial cross sections near the $E=0$ threshold approach 0 with a slope
that depends on $a_s^2$.

\begin{figure}
\vspace*{0.5cm}
\hspace*{0cm}
\includegraphics[width=0.45\textwidth]{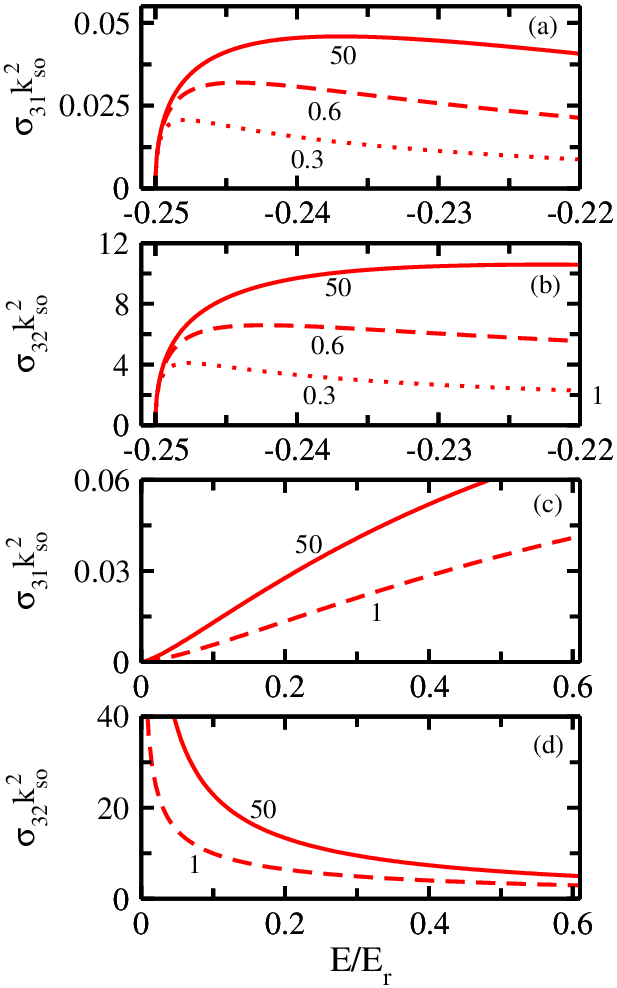}
\caption{(Color online)
Partial cross sections,
in units of $1/k_{\text{so}}^2$,  
for the $(J,M_J)=(1,M_J)$ channels as a function of the energy $E$
for (a) and (b) unequal masses 
$(\eta=1/2)$, 
and 
(c) and (d) equal masses ($\eta=0$);
this figure considers scattering processes from branch B to branch A.
Panel~(a) shows 
$\sigma_{31}k_{\text{so}}^2=\sigma_{41}k_{\text{so}}^2$ 
in the vicinity of $E=-\eta^2E_r$
for $|a_sk_{\text{so}}|=50$
(solid line), 
$|a_sk_{\text{so}}|=0.6$ (dashed line), and $|a_sk_{\text{so}}|=0.3$ (dotted line).
Panel~(b) shows 
$\sigma_{32}k_{\text{so}}^2=\sigma_{42}k_{\text{so}}^2$ 
in the vicinity of $E=-\eta^2E_r$
for $|a_sk_{\text{so}}|=50$
(solid line), 
$|a_sk_{\text{so}}|=0.6$ (dashed line), and $|a_sk_{\text{so}}|=0.3$ (dotted line).
Panel~(c) shows 
$\sigma_{31}k_{\text{so}}^2=\sigma_{41}k_{\text{so}}^2$ 
in the vicinity of $E=0$
for $|a_sk_{\text{so}}|=50$
(solid line) and 
$|a_sk_{\text{so}}|=1$ (dashed line).
Panel~(d) shows 
$\sigma_{32}k_{\text{so}}^2=\sigma_{42}k_{\text{so}}^2$ 
in the vicinity of $E=0$
for $|a_sk_{\text{so}}|=50$
(solid line) and 
$|a_sk_{\text{so}}|=1$ (dashed line). 
}
\label{fig6}
\end{figure}

Finally, we analyze the partial cross 
sections $\sigma_{31}$, $\sigma_{41}$, $\sigma_{32}$ and $\sigma_{42}$,
which describe scattering processes from
states in branch B to states in branch A.
Taylor expanding $\sigma_{31}=\sigma_{41}$ and $\sigma_{32}=\sigma_{42}$
around $E=-\eta^2E_r$ ($E \ge -\eta^2 E_r$),
we find that the leading order terms
of these partial cross sections
are, for $\eta>0$, independent of $a_s$ and proportional to 
$\sqrt{|E+\eta^2E_r|}$
(see Table~\ref{table:threshold_1M_J_31_41}).
\begin{table}
\begin{center}
\begin{tabular}{c|c|c}
  &$E\to(-\eta^2E_r)^{+}$ &$E\to 0^+$\\
  &$(\eta, a_s \ne 0)$ &
$(\eta=0,a_s \ne 0)$\\
  \hline
  $\sigma_{31}k_{\text{so}}^2=\sigma_{41}k_{\text{so}}^2$
  &$c_{31}^{(0)}y$ &$d_{31}^{(0)}y^3$\\
  $\sigma_{32}k_{\text{so}}^2=\sigma_{42}k_{\text{so}}^2$
  &$c_{32}^{(0)}y$ &$d_{32}^{(0)}y^{-1}$
\end{tabular}
\caption{Behavior of the partial cross sections 
for two particles
with unequal masses ($\eta\ne 0$) and equal masses ($\eta=0$) in the $(J,M_J)=(1,M_J)$ channels
in the vicinity of the energy where branch B becomes closed;
this table considers scattering processes from branch B to branch A.
The coefficients $c_{31}^{(0)}$ and $c_{32}^{(0)}$
for $\eta\ne 0$,
which are reported in Appendix~\ref{appendixB},
are independent of $a_s$.
In contrast,
the coefficients $d_{31}^{(0)}$ and $d_{32}^{(0)}$
for $\eta=0$, which are also reported in Appendix~\ref{appendixB},
depend on $a_s^2$.
}
\label{table:threshold_1M_J_31_41}
\end{center}
\end{table}
Figures~\ref{fig6}(a) and \ref{fig6}(b)
illustrate the threshold behaviors
for $\sigma_{31}$ and $\sigma_{32}$, respectively,
in the vicinity of $E=-\eta^2 E_r$
for $\eta=1/2$ and $|a_sk_{\text{so}}|=50$, $0.6$, and $0.3$.
Figures~\ref{fig6}(c) and \ref{fig6}(d) show $\sigma_{31}$ and $\sigma_{32}$,
respectively,
for $\eta=0$ and $|a_sk_{\text{so}}|=50$ and $1$.
For $\eta=0$,
$\sigma_{31}=\sigma_{41}$ goes to zero at the 
$E=0$ 
threshold, with the 
leading order term depending on $a_s^2$,
while $\sigma_{32}=\sigma_{42}$ diverges as $1/\sqrt{|E|}$,
with the 
leading order term depending on $a_s^2$.
We note that an analogous
divergent behavior of a subset of the partial cross sections
has been predicted for a spin-1 system with spin-orbit coupling~\cite{chris}.

As in Sec.~\ref{sec3},
we 
diagonalize the $\mathcal{K}^{(1,M_J)}$ matrix for $E\ge -\eta^2E_r$ 
and the 
$\mathcal{K}^{(1,M_J)}_{oo}$ 
matrix for $E<-\eta^2E_r$, 
and transform to the
rotated or
standing wave basis.
For $E\ge -\eta^2E_r$,
we find that three 
of the
eigenvalues of 
$\mathcal{K}^{(1,M_J)}$ are zero.
If we had used a pseudo-potential that accounts not only
for $s$-wave interactions but also for higher-partial
wave interactions, these eigenvalues would not be zero.
The fourth eigenvalue reads
\begin{eqnarray}
\label{phase_shift_2}
\text{tan}\delta_{\text{eff,1}}^s=
-\frac{a_sk_{\text{so}}}{3}
\Bigg[
\left(\frac{E}{E_r}+1\right)^{1/2}+\left(\frac{E}{E_r}+1\right)^{-1/2}+ \nonumber \\
2\left(\frac{E}{E_r}+\eta^2\right)^{1/2}+2\eta^2\left(\frac{E}{E_r}+\eta^2\right)^{-1/2}\Bigg].
\end{eqnarray}
Equation~\eqref{phase_shift_2} shows 
that
the threshold behavior around $-\eta^2E_r$ depends 
strongly on $\eta$.
For $\eta = 0$, 
the last term in square bracket in Eq.~\eqref{phase_shift_2}
vanishes and 
$\tan\delta_{\text{eff},1}^s$ behaves like 
$-2a_sk_{\text{so}}(1+\sqrt{E/E_r})/3$ in the vicinity of $E=0$.
This implies that the 
partial cross sections depend
on $a_s$.
For $\eta\ne 0$,
the last term in square bracket in Eq.~\eqref{phase_shift_2},
which behaves like a one-dimensional system, dominates (in fact, it diverges).
Correspondingly, 
the partial cross sections 
are
independent 
of $a_s$ as 
$E \rightarrow (-\eta^2E_r)^+$.

For $E<-\eta^2E_r$, 
we find that one 
of the eigenvalues
of $\mathcal{K}^{(1,M_J)}_{oo}$ is zero while 
the other reads
\begin{eqnarray}
\label{phase_shift_3}
\text{tan}\delta_{\text{eff,2}}^s
=-a_{\text{eff}}k_{\text{so}}
\left[\left(\frac{E}{E_r}+1\right)^{1/2}+\left(\frac{E}{E_r}+1\right)^{-1/2}\right].
\end{eqnarray}
Equation~\eqref{phase_shift_3} shows that near the lowest scattering
threshold 
$(E=-E_r)$,
the tangent of the 
phase shift is approximately proportional to
the density of states of a one-dimensional system
for both $\eta\ne 0$ and $\eta = 0$ 
(i.e., the second term in the square bracket dominates as $E\to -E_r$).
This
is similar to 
what we found in Sec.~\ref{sec3}
for
the $(J,M_J)=(0,0)$ channel.
For $E=-E_r$,
the tangent of the 
phase shift diverges and the 
partial cross sections
are independent of $a_s$.

This section discussed the scattering processes
within branch A (see Figs.~\ref{fig3a} and \ref{fig3b} and Table~\ref{threshold_11_12}),
within branch B (see Fig.~\ref{fig4} and Table~\ref{table:threshold_1M_J_33_34}),
from branch A to branch B (see Fig.~\ref{fig5} and Table~\ref{table:threshold_1M_J_13_14}),
and
from branch B to branch A (see Fig.~\ref{fig6} and Table~\ref{table:threshold_1M_J_31_41})
for the $(J,M_J)=(1,M_J)$ channels.
As a means of summarizing, we 
consider the limiting values of the
cross section matrices at the scattering thresholds.
At the lowest scattering threshold ($E=-E_r$), the cross section
matrix approaches, regardless of the mass ratio,
\begin{eqnarray}
\label{thres_matrix_1}
\sigma\xrightarrow{E=-E_r}
\left[
\begin{array}{cc}
\frac{2\pi}{k_{\text{so}}^2} &\frac{2\pi}{k_{\text{so}}^2} \\
\frac{2\pi}{k_{\text{so}}^2} &\frac{2\pi}{k_{\text{so}}^2}\\
\end{array}
\right].
\end{eqnarray}  
For equal masses, the behavior of $\sigma_{12}$
is shown in Figs.~\ref{fig3a}(b) and \ref{fig3b}(d).
For unequal masses, the behavior of $\sigma_{12}$
is shown in Figs.~\ref{fig3a}(a) and \ref{fig3b}(b). 
As already discussed, the mass ratio
provides a means to tune
the scattering threshold at which branch B becomes closed
and, subsequently, to tune the threshold behavior.
For equal masses, branch B becomes closed at $E=0$ and we have
\begin{eqnarray}
\label{thres_matrix_2}
\sigma\xrightarrow{E=0^+}
\begin{bmatrix}
 0^{\protect\ref{fig3b}\text{(c)}} &\frac{d_{12}^{(0)}(a_s)}{k_{\text{so}}^2}^{\protect\ref{fig3a}\text{(b)}} &0^{\protect\ref{fig5}\text{(c)}} &0^{\protect\ref{fig5}\text{(c)}}\\
 0^{\protect\ref{fig3b}\text{(c)}} &\frac{d_{12}^{(0)}(a_s)}{k_{\text{so}}^2}^{\protect\ref{fig3a}\text{(b)}} &0^{\protect\ref{fig5}\text{(c)}} &0^{\protect\ref{fig5}\text{(c)}}\\
 0^{\protect\ref{fig6}\text{(c)}} &\infty^{\protect\ref{fig6}\text{(d)}} &\frac{d_{12}^{(0)}(a_s)}{k_{\text{so}}^2}^{\protect\ref{fig4}\text{(c)}} &\frac{d_{12}^{(0)}(a_s)}{k_{\text{so}}^2}^{\protect\ref{fig4}\text{(c)}}\\
 0^{\protect\ref{fig6}\text{(c)}} &\infty^{\protect\ref{fig6}\text{(d)}} &\frac{d_{12}^{(0)}(a_s)}{k_{\text{so}}^2}^{\protect\ref{fig4}\text{(c)}} &\frac{d_{12}^{(0)}(a_s)}{k_{\text{so}}^2}^{\protect\ref{fig4}\text{(c)}}\\
\end{bmatrix}.
\end{eqnarray}
For unequal masses, branch B becomes closed at $E=-\eta^2 E_r$ and we instead have
\begin{eqnarray}
\label{thres_matrix_3}
\sigma\xrightarrow{E=-\eta^2E_r}
\begin{bmatrix}
 0^{\protect\ref{fig3b}\text{(a)}} &0^{\protect\ref{fig3a}\text{(a)}} &0^{\protect\ref{fig5}\text{(a)}} &0^{\protect\ref{fig5}\text{(b)}}\\
 0^{\protect\ref{fig3b}\text{(a)}} &0^{\protect\ref{fig3a}\text{(a)}} &0^{\protect\ref{fig5}\text{(a)}} &0^{\protect\ref{fig5}\text{(b)}}\\
 0^{\protect\ref{fig6}\text{(a)}} &0^{\protect\ref{fig6}\text{(b)}} &\frac{2\pi}{\eta^2k_{\text{so}}^2}^{\protect\ref{fig4}\text{(a)}} &\frac{2\pi}{\eta^2k_{\text{so}}^2}^{\protect\ref{fig4}\text{(b)}}\\
 0^{\protect\ref{fig6}\text{(a)}} &0^{\protect\ref{fig6}\text{(b)}} &\frac{2\pi}{\eta^2k_{\text{so}}^2}^{\protect\ref{fig4}\text{(a)}} &\frac{2\pi}{\eta^2k_{\text{so}}^2}^{\protect\ref{fig4}\text{(b)}}\\
\end{bmatrix}.
\end{eqnarray}  
In Eqs.~(\ref{thres_matrix_2}) and (\ref{thres_matrix_3}), the superscript 
indicates the figure number in which the corresponding cross section is shown
[e.g., the entry ``$0^{\protect\ref{fig3b}\text{(a)}}$'' 
in Eq.~(\ref{thres_matrix_3}) says that
$\sigma_{11}$ is equal to zero at $E=-\eta^2E_r$ and that this behavior can
be seen in Fig.~\ref{fig3b}(a)].
Comparison of Eqs.~(\ref{thres_matrix_2}) and (\ref{thres_matrix_3}) 
shows that 
the threshold behavior
is significantly impacted by
the mass ratio.

\begin{figure}
\vspace*{1.5cm}
\hspace*{0cm}
\includegraphics[width=0.4\textwidth]{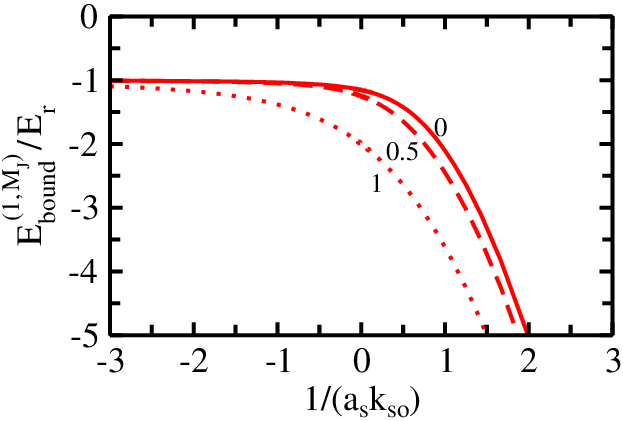}
\caption{(Color online)
Bound state energy $E_{\text{bound}}^{(1,M_J)}$ for 
two particles in the $(J,M_J)=(1,M_J)$ channel.
Solid, dashed and dotted lines show
$E_{\text{bound}}^{(1,M_J)}$ for 
$\eta=0$, $1/2$ and $1$, respectively.
For increasing $\eta$, $E_{\text{bound}}^{(1,M_J)}$ decreases, i.e., the dimer becomes---if
the binding energy is measured in units of $E_r$---more strongly bound
with increasing mass imbalance.
In the extreme $\eta=1$ case,
$E_{\text{bound}}^{(1,M_J)}$ is identical to $E_{\text{bound}}^{(0,0)}$.
}
\label{fig7}
\end{figure}

To determine the bound state energies $E_{\text{bound}}^{(1,M_J)}$
for the $(J,M_J)=(1,M_J)$ channels,  
we analyze the poles of the $\mathcal{S}^{(1,M_J)}$ matrix.
The 
bound state
energy 
$E_{\text{bound}}^{(1,M_J)}$ is determined by the roots of the equation
\begin{eqnarray}
  \label{equation_bound_energy}
\tan \delta_{\text{eff,1}}^s=-\imath.
\end{eqnarray}
Since $\mathcal{S}^{(1,M_J)}_{oo}$ is 
a
sub-matrix of $\mathcal{S}^{(1,M_J)}$,
the poles of $\mathcal{S}^{(1,M_J)}_{oo}$ are 
also
poles of $\mathcal{S}^{(1,M_J)}$.
The equation $\tan\delta_{\text{eff,2}}^s=-\imath$ 
yields the same result for the bound state energy. 
In the weak-binding limit, i.e., for $1/a_s\to -\infty$,
the bound state energy approaches $-(1+a_s^2k_{\text{so}}^2/9)E_r$.
In the strong-binding limit, i.e., for $1/a_s \to +\infty$,
the bound state energy approaches $-\hbar^2/(2\mu a_s^2)-(1+2\eta^2)E_r$.
The 
solid, dashed and dotted lines in Fig.~\ref{fig7} show 
$E_{\text{bound}}^{(1,M_J)}/E_r$,
$E_{\text{bound}}^{(1,M_J)} < -E_r$,
 as a function of $1/(a_sk_{\text{so}})$ for $\eta=0$, $1/2$ and $1$,
respectively.
For all $\eta$, there exists 
a single
two-body bound state for all $a_s$.
For fixed $a_sk_{\text{so}}$, $E_{\text{bound}}^{(1,M_J)}$ becomes more negative as $\eta$
increases from $0$ to $1$.
For $\eta=1$, 
$E_{\text{bound}}^{(1,M_J)}$ 
is identical to
$E_{\text{bound}}^{(0,0)}$
since $\tan\delta_{\text{eff},1}^s$, Eq.~\eqref{phase_shift_2}, 
agrees with $\tan\delta_{\text{eff}}^s$, Eq.~\eqref{phase_shift},
in this case.

\section{conclusion}
\label{sec5}
This paper
developed---building on the formulations presented in
Refs.~\cite{Gao,chris} for positive energies---a 
scattering framework for two particles with 
isotropic spin-orbit coupling applicable to all energies.
Inspired by the 
usual partial wave decomposition 
for systems without spin-orbit coupling,
Refs.~\cite{Gao,chris} derived solutions
using the magnitude of the wave number $k$.
Since 
the derivative of the single particle energies
with respect to $k$ 
for systems with isotropic spin-orbit coupling
shows a discontinuity at vanishing energy [see Fig.~\ref{fig1}(a)],
the extension of the framework developed in Refs.~\cite{Gao,chris}
to negative energies is not entirely straightforward.
One approach would be to obtain solutions for the various energy regions
separately and to then match the solutions.
Alternatively, and this
is the route pursued 
in this paper, one might seek solutions
that apply to all energy regions.    
To this end, we replaced $k$ by $\bar{k}$, $\bar{k}=h_{\text{rel}} k$,
and defined single-particle energies whose derivative with respect
to 
$\bar{k}$
is continuous for all energies [see Figs.~\ref{fig1}(b) and \ref{fig1}(c)].
Our formulation allows us to treat the
various scattering thresholds and energy
variations across these thresholds using standard
coupled-channel formulations. 
Specifically, this opens the door for calculating energy-dependent
or thermally averaged scattering cross sections by making 
modifications to 
existing
propagation based scattering codes.

We applied
our framework to the $(J,M_J)=(0,0)$ and $(J,M_J)=(1,M_J)$
channels for two particles, with equal or unequal masses, interacting
through the zero-range $s$-wave pseudo-potential. 
Assuming $s$-wave contact interactions,
the $J=0$ and $1$ channels are the only channels affected 
by the two-body interactions,
i.e.,
the behavior of the $J=2,3,...$ channels 
is independent of $a_s$.
If we, e.g.,
take the state $|\bar{k},\theta_{\mathbf{k}},\phi_{\mathbf{k}}\rangle_A$
as our initial state,
the population of the $J=1$ partial wave channels is 
three times higher than that of the $J=0$ partial wave channel.
If we take an equal mixture of 
$|\bar{k},\theta_{\mathbf{k}},\phi_{\mathbf{k}}\rangle_A$
and
$|\bar{k},\theta_{\mathbf{k}},\phi_{\mathbf{k}}\rangle_B$
as our initial state,
then the population of the $J=1$ partial wave channels is 
six times higher than that of the $J=0$ partial wave channel.
While this paper did not report total cross sections,
the above statements give a feeling for 
how to convert the partial wave cross sections presented in this paper to total cross sections.
  
We obtained closed analytical expressions 
applicable to all energies for the partial wave cross sections.
The behavior of the partial cross sections 
was 
analyzed in detail near the 
scattering thresholds. Particular attention was paid
to the 
scattering thresholds located at
negative energies and simple analytical results
were reported for the  limiting behaviors.
While our results were obtained for a specific
functional form of the spin-orbit coupling and specific 
scattering channels, we believe that our study points
toward more general characteristics of
two-body scattering in the presence of spin-orbit coupling.
First,
the partial cross sections for scattering between 
states corresponding to a branch that becomes closed at 
a particular 
negative energy
scattering threshold are---at the scattering 
threshold---independent
of the $s$-wave scattering length
$a_s$ and fully determined by the
spin-orbit coupling strength $k_{\text{so}}$.
This universal behavior can be interpreted as being a consequence
of an effective dimensionality reduction near
the scattering thresholds. At the lowest scattering threshold,
all partial wave cross sections, and hence also the total cross section, are independent
of $a_s$.
Second,
the mass ratio dependence of the results in the
$(J,M_J)=(1,M_J)$ channels points toward an interesting
tunability of few- and many-body properties of mass 
imbalanced systems. 
It is easily recognized that the scattering threshold where 
branch B becomes closed depends on the mass ratio. Since 
a subset of the partial cross sections 
vanishes
at this
threshold, different mass ratios should result
in different energy dependent cross sections.
Moreover, the two-body binding energy was found to depend 
on the mass ratio.

The present paper suggests a variety of follow-up studies.
Most immediately, it would be interesting to extend our 
theoretical framework to systems with finite total
momentum or with other types of spin-orbit coupling.
Moreover, it would be interesting to incorporate the
formulation presented in this paper into
a two-body scattering code and to investigate the
influence of two-body van der Waals physics on 
scattering observables in the presence of
spin-orbit coupling.
Looking further ahead, extending
the scattering framework to
three particles with spin-orbit coupling would be a major step forward.

\section{acknowledgement}
\label{acknowledgement}
Support by the National Science Foundation through
grant number  
PHY-1509892,
discussions with X. Cui and Y. Yan,
and comments on the manuscript by J. Jacob 
are gratefully acknowledged.

\appendix
\section{Coefficients that describe 
the
threshold behavior  
of
the $(J,M_J)=(1,M_J)$ channel}
\label{appendixB}
In the following we report explicit
expressions for
the coefficients 
introduced
in Table~\ref{threshold_11_12}:
\begin{eqnarray}
c_{11}^{(0)} & =& \frac{\pi}{2(1-\eta^2)(1+\sqrt{1-\eta^2})^2},
\end{eqnarray}
\begin{eqnarray}
c_{11}^{(1+)} & =& \frac{\eta^2-2}{\eta^2(1-\eta^2)^{1/2}} c_{11}^{(0)},
\end{eqnarray}
\begin{eqnarray}
c_{11}^{(1-)} & =& \frac{3}{\eta^2}c_{11}^{(0)},
\end{eqnarray}
\begin{eqnarray}
c_{12}^{(0)} & =& \frac{\pi}{2(1-\eta^2)(1-\sqrt{1-\eta^2})^2},
\end{eqnarray}
\begin{eqnarray}
c_{12}^{(1+)} & =& \frac{\eta^2-2}{\eta^2(1-\eta^2)^{1/2}} c_{12}^{(0)},
\end{eqnarray}
\begin{eqnarray}
c_{12}^{(1-)} & =& \frac{3}{\eta^2}c_{12}^{(0)},
\end{eqnarray}

\begin{eqnarray}
d_{11}^{(0)} & =& \frac{\pi a_s^2k_{\text{so}}^2}{2(9+4a_s^2k_{\text{so}}^2)},
\end{eqnarray}
\begin{eqnarray}
d_{11}^{(1)} & =& -\frac{16}{\pi}(d_{11}^{(0)})^2,
\end{eqnarray}
\begin{eqnarray}
d_{12}^{(0)} & =& 16d_{11}^{(0)},
\end{eqnarray}
and
\begin{eqnarray}
d_{12}^{(1)}  = -\frac{1}{\pi}(d_{12}^{(0)})^2.
\end{eqnarray}
In the following we report explicit
expressions for
the coefficients 
introduced
in Table~\ref{table:threshold_1M_J_13_14}:
\begin{eqnarray}
c_{13}^{(0)}=c_{14}^{(0)} & =& \frac{\pi}{\eta^2\sqrt{1-\eta^2}}
\end{eqnarray}
and
\begin{eqnarray}
d_{13}^{(0)}=d_{14}^{(0)} & =& \frac{4\pi a_s^2k_{\text{so}}^2}{9+4a_s^2k_{\text{so}}^2}.
\end{eqnarray}
In the following we report explicit
expressions for
the coefficients 
introduced
in Table~\ref{table:threshold_1M_J_31_41}:
\begin{eqnarray}
c_{31}^{(0)} & =& 2\sqrt{1-\eta^2}c_{11}^{(0)},\\
c_{32}^{(0)} & =& 2\sqrt{1-\eta^2}c_{12}^{(0)},\\
d_{31}^{(0)} & =& \frac{1}{4}d_{13}^{(0)},
\end{eqnarray}
and
\begin{eqnarray}
d_{32}^{(0)} & =& 4d_{13}^{(0)}.
\end{eqnarray}


\begin{thebibliography}{100}

\bibitem{condense_matter}
G. Dresselhaus,
Spin-Orbit Coupling Effects in Zinc Blende Structures,
Phys. Rev. {\bf{100}}, 580 (1955).

\bibitem{condense_matter1}
Yu. A. Bychkov and E. I. Rashba, 
Properties of a 2D electron gas with a lifted spectrum degeneracy, 
Sov. Phys.- JETP Lett. {\bf{39}}, 78 (1984).

\bibitem{sengstock}
J. Struck,
C. \"Olschl\"ager, M. Weinberg, P. Hauke, J. Simonet, A. Eckardt, M. Lewenstein, K. Sengstock, and P. Windpassinger,
Tunable Gauge Potential for Neutral and Spinless Particles in Driven Optical Lattices,
Phys. Rev. Lett. {\bf{108}}, 225304 (2012).

\bibitem{spielman}
Y.-J. Lin, K. Jim\'enez-Garc{\'\i}a, and I. B. Spielman, 
Spin-orbit-coupled Bose-Einstein condensates, 
Nature (London) {\bf{471}}, 83 (2011).

\bibitem{zwierlein}
L. W. Cheuk, A. T. Sommer, Z. Hadzibabic, T. Yefsah, W. S. Bakr, and M. W. Zwierlein,
Spin-Injection Spectroscopy of a Spin-Orbit Coupled Fermi Gas,
Phys. Rev. Lett. {\bf{109}}, 095302 (2012).

\bibitem{zhang}
P. Wang, Z.-Q. Yu, Z. Fu, J. Miao, L. Huang, S. Chai, H. Zhai, and J. Zhang,
Spin-Orbit Coupled Degenerate Fermi Gases,
Phys. Rev. Lett. {\bf{109}}, 095301 (2012).

\bibitem{pan}
J.-Y. Zhang, S.-C. Ji, Z. Chen, L. Zhang, Z.-D. Du, B. Yan, G.-S. Pan, B. Zhao, Y.-J. Deng, H. Zhai, S. Chen, and J.-W. Pan,
Collective Dipole Oscillations of a Spin-Orbit Coupled Bose-Einstein Condensate,
Phys. Rev. Lett. {\bf{109}}, 115301 (2012). 

\bibitem{2D_SOC}
L. Huang, Z. Meng, P. Wang, P. Peng, S.-L. Zhang, L. Chen, D. Li, Q. Zhou, and J. Zhang,
Experimental realization of two-dimensional synthetic spin-orbit coupling in ultracold Fermi gases,
Nat. Phys. {\bf{12}}, 540 (2016).

\bibitem{proposal_1}
J. Dalibard, F. Gerbier, G. Juzeli\={u}nas, and P. \"Ohberg,
Colloquium: Artificial gauge potentials for neutral atoms,
Rev. Mod. Phys. {\bf{83}}, 1523 (2011).

\bibitem{proposal_2}
B. M. Anderson, G. Juzeli\={u}nas, V. M. Galitski, and I. B. Spielman,
Synthetic 3D Spin-Orbit Coupling,
Phys. Rev. Lett. {\bf{108}}, 235301 (2012).

\bibitem{proposal_3}
B. M. Anderson, I. B. Spielman, and G. Juzeli\={u}nas,
Magnetically Generated Spin-Orbit Coupling for Ultracold Atoms,
Phys. Rev. Lett. {\bf{111}}, 125301 (2013).

\bibitem{rashba_billiard}
J. Cserti, A. Csord\'as, and U. Z\"ulicke,
Electronic and spin properties of Rashba billards,
Phys. Rev. B {\bf{70}}, 233307 (2004).

\bibitem{Novikov}
D. S. Novikov, 
Elastic scattering theory and transport in graphene,
Phys. Rev. B {\bf{76}}, 245435 (2007).

\bibitem{Joel}
J. Hutchinson and J. Maciejko,
Rashba scattering in the low-energy limit,
Phys. Rev. B {\bf{93}}, 245309 (2016).

\bibitem{landau_zener}
A. J. Olson, S.-J. Wang, R. J. Niffenegger, C.-H. Li, C. H. Greene, and Y. P. Chen,
Tunable Landau-Zener transitions in a spin-orbit-coupled Bose-Einstein condensate,
Phys. Rev. A {\bf{90}}, 013616 (2014).

\bibitem{Zitterbewegung}
C. Qu, C. Hamner, M. Gong, C. Zhang, and P. Engels,
Observation of Zitterbewegung in a spin-orbit-coupled Bose-Einstein condensate,
Phys. Rev. A {\bf{88}}, 021604(R) (2013).

\bibitem{spin_wave}
Y. Li, C. Qu, Y. Zhang, and C. Zhang,
Dynamical spin-density waves in a spin-orbit-coupled Bose-Einstein condensate,
Phys. Rev. A {\bf{92}}, 013635 (2015).

\bibitem{ming_gong}
M. Gong, S. Tewari, and C. Zhang,
BCS-BEC Crossover and Topological Phase Transition in 3D Spin-Orbit Coupled Degenerate Fermi Gases,
Phys. Rev. Lett. {\bf{107}}, 195303 (2011).

\bibitem{H_Hu}
H. Hu, L. Jiang, X.-J Liu, and H. Pu,
Probing Anisotropic Superfluidity in Atomic Fermi Gases with Rashba Spin-Orbit Coupling,
Phys. Rev. Lett. {\bf{107}}, 195304 (2011).

\bibitem{crossover_hui}
Z.-Q. Yu and H. Zhai,
Spin-Orbit Coupled Fermi Gases across a Feshbach Resonance,
Phys. Rev. Lett. {\bf{107}}, 195305 (2011).

\bibitem{bound_shenoy}
J. P. Vyasanakere and V. B. Shenoy,
Bound states of two spin-1/2 fermions in a synthetic non-Abelian gauge field,
Phys. Rev. B {\bf{83}}, 094515 (2011).

\bibitem{crossover_shenoy}
J. P. Vyasanakere, S. Zhang, and V. B. Shenoy,
BCS-BEC crossover induced by a synthetic non-Abelian gauge field,
Phys. Rev. B {\bf{84}}, 014512 (2011).

\bibitem{rashbon_shenoy}
J. P. Vyasanakere and V. B. Shenoy,
Rashbons: properties and their significance,
New. J. Phys. {\bf{14}}, 043041 (2012).

\bibitem{L_Han}
L. Han and C. A. R. S\'a de Melo,
Evolution from BCS to BEC superfluidity in the presence of spin-orbit coupling,
Phys. Rev. A {\bf{85}}, 011606(R) (2012).

\bibitem{zhang_jing}
Z. Fu, L. Huang, Z. Meng, P. Wang, L. Zhang, S. Zhang, H. Zhai, P. Zhang, and J. Zhang,
Production of Feshbach molecules induced by spin-orbit coupling in Fermi gases,
Nat. Phys. {\bf{10}}, 110 (2014).

\bibitem{scattering_experiment}
R. A. Williams, L. J. LeBlanc, K. Jim\'enez-Garc{\'\i}a, M. C. Beeler, A. R. Perry, W. D. Phillips, and I. B. Spielman,
Synthetic Partial Waves in Ultracold Atomic Collisions,
Science {\bf{335}}, 314 (2012). 

\bibitem{stripe_hui}
C. Wang, C. Gao, C.-M. Jian, and H. Zhai,
Spin-Orbit Coupled Spinor Bose-Einstein Condensates,
Phys. Rev. Lett. {\bf{105}}, 160403 (2010).

\bibitem{jason}
T.-L. Ho and S. Zhang,
Bose-Einstein Condensates with Spin-Orbit Interaction,
Phys. Rev. Lett. {\bf{107}}, 150403 (2011).

\bibitem{stripe}
Y. Zhang, L. Mao, and C. Zhang,
Mean-Field Dynamics of Spin-Orbit Coupled Bose-Einstein Condensates,
Phys. Rev. Lett. {\bf{108}}, 035302 (2012).

\bibitem{stringari}
Y. Li, G. I. Martone, L. P. Pitaevskii, and S. Stringari,
Superstripes and the Excitation Spectrum of a Spin-Orbit-Coupled Bose-Einstein Condensate,
Phys. Rev. Lett. {\bf{110}}, 235302 (2013).


\bibitem{spin_chain_cui}
X. Cui and T.-L. Ho,
Spin-orbit-coupled one-dimensional Fermi gases with infinite repulsion,
Phys. Rev. A {\bf{89}}, 013629 (2014).

\bibitem{spin_chain_guan}
Q. Guan and D. Blume,
Spin structure of harmonically trapped one-dimensional atoms with spin-orbit coupling,
Phys. Rev. A {\bf{92}}, 023641 (2015).

\bibitem{three_body_hui}
Z.-Y. Shi, X. Cui, and H. Zhai,
Universal Trimers Induced by Spin-Orbit Coupling in Ultracold Fermi Gases,
Phys. Rev. Lett. {\bf{112}}, 013201 (2014).

\bibitem{three_body_yi}
X. Cui and W. Yi,
Universal Borromean Binding in Spin-Orbit-Coupled Ultracold Fermi Gases,
Phys. Rev. X {\bf{4}}, 031026 (2014).

\bibitem{three_body_cui}
Z.-Y. Shi, H. Zhai, and X. Cui,
Efimov physics and universal trimers in spin-orbit-coupled ultracold atomic mixtures,
Phys. Rev. A {\bf{91}}, 023618 (2015).

\bibitem{Gao}
H. Duan, L. You, and B. Gao, 
Ultracold collisions in the presence of synthetic spin-orbit coupling,
Phys. Rev. A {\bf{87}}, 052708 (2013).

\bibitem{chris}
S.-J. Wang and C. H. Greene, 
General formalism for ultracold scattering with isotropic spin-orbit coupling, 
Phys. Rev. A {\bf{91}}, 022706 (2015).

\bibitem{pengzhang}
L. Zhang, Y. Deng, and P. Zhang,
Scattering and effective interactions of ultracold atoms with spin-orbit coupling, 
Phys. Rev. A {\bf{87}}, 053626 (2013).

\bibitem{pengzhang1}
P. Zhang, L. Zhang, and Y. Deng,
Modified Bethe-Peierls boundary condition for ultracold atoms with spin-orbit coupling,
Phys. Rev. A {\bf{86}}, 053608 (2012).

\bibitem{xiaoling}
X. Cui, 
Mixed-partial-wave scattering with spin-orbit coupling and validity of pseudopotentials,
Phys. Rev. A {\bf{85}}, 022705 (2012).

\bibitem{zhenhua}
Y. Wu and Z. Yu,
Short-range asymptotic behavior of the wave functions of interacting spin-1/2 fermionic atoms with spin-orbit coupling: A model study,
Phys. Rev. A {\bf{87}}, 032703 (2013).

\bibitem{review_hui}
H. Zhai,
Degenerate quantum gases with spin-orbit coupling: a review,
Rep. Prog. Phys. {\bf{78}}, 026001 (2015).

\bibitem{comment_on_permute}
There are three separate commutators to consider for the
$x$-, $y$- and $z$-components of $\hat{\mathbf{P}}$.
Moreover, since $\hat{H}_{\text{tot}}$ is a $4$ by $4$ 
matrix, the commutator $[\hat{H}_{\text{tot}},\hat{P}_x]$ can be
broken down
into 16 separate commutators (and similarly for
$\hat{P}_y$ and $\hat{P}_z$).

\bibitem{expectation_value_argument}
Since the expectation value is 
taken
with respect to 
an
eigenstate of $\hat{\mathbf{p}}$,
we have 
$|\langle\hat{\mathbf{p}}\cdot\hat{\mathbf{\Sigma}}\rangle|=|\langle\hat{\mathbf{p}}\rangle\cdot\langle\hat{\mathbf{\Sigma}}\rangle|=p|\langle\hat{\mathbf{\Sigma}}\rangle|=\hbar k |\langle\hat{\mathbf{\Sigma}}\rangle|$.

\bibitem{zare}
R. N. Zare, {\it{Angular Momentum: Understanding Spatial Aspects in Chemistry and Physics}}, 1st ed. (Wiley-Interscience, 1991).

\bibitem{note_on_normalization}
Our current normalization condition differs from the Wronskian based normalization condition 
in Ref.~\cite{chris}; 
as a result, the unit of our normalization constants $N$ differs from that of Ref.~\cite{chris}.

\bibitem{partial_cross_section_argument}
We refer to the $\sigma_{jl}^{(J,M_J)}$ as ``partial cross sections''
since various $(J,M_J)$ combinations contribute to the scattering from channel $j$
to channel $l$.
Reference~\cite{Gao} used instead the term ``cross section'' and Ref.~\cite{chris} the term ``(total) cross section''. 

\bibitem{chin}
C. Chin, R. Grimm, P. Julienne, and E. Tiesinga,
Feshbach resonances in ultracold gases,
Rev. Mod. Phys. {\bf{82}}, 1225 (2010). 

\bibitem{tune_soc}
Y. Zhang, G. Chen, and C. Zhang,
Tunable Spin-orbit Coupling and Quantum Phase Transition in a Trapped Bose-Einstein Condensate,
Sci. Rep. {\bf{3}}, 1937 (2013).

\bibitem{tune_soc2}
K. Jim\'enez-Garc\'ia, L. J. LeBlanc, R. A. Williams, M. C. Beeler, C. Qu, M. Gong, C. Zhang, and I. B. Spielman,
Tunable Spin-Orbit Coupling via Strong Driving in Ultracold-Atom Systems,
Phys. Rev. Lett. {\bf{114}}, 125301 (2015).

\bibitem{tune_soc3}
X. Luo, L. Wu, J. Chen, Q. Guan, K. Gao, Z.-F Xu, L. You, and R. Wang,
Tunable atomic spin-orbit coupling synthesized with a modulating gradient magnetic field,
Sci. Rep. {\bf{6}}, 18983 (2016).

\bibitem{johnson}
B. R. Johnson,
Comment on a recent criticism of the formula used to calculate the $S$ matrix in the multichannel log-derivative method,
Phys. Rev. A {\bf{32}}, 1241 (1985).

\end{thebibliography}
\end{document}